\documentclass[11pt,epsf]{article}

\usepackage{amsmath}
\usepackage{graphicx}
\usepackage{indentfirst}
\usepackage{cite}
\usepackage{color}
\usepackage{subfigure}
\usepackage[titletoc,title]{appendix}

\begin{document}

\title{\bf{Collision of Two Rotating Hayward Black Holes}}

\date{}
\maketitle

\begin{center}
\author{Bogeun Gwak}$^a$\footnote{rasenis@sejong.ac.kr}\\

\vskip 0.25in
$^{a}$\it{Department of Physics and Astronomy, Sejong University, Seoul 05006, Republic of Korea}\\
\end{center}
\vskip 0.6in

{{\abstract
We investigate the spin interaction and the gravitational radiation thermally allowed in a head-on collision of two rotating Hayward black holes. The Hayward black hole is a regular black hole in a modified Einstein equation, and hence it can be an appropriate model to describe the extent to which the regularity effect in the near-horizon region affects the interaction and the radiation. If one black hole is assumed to be considerably smaller than the other, the potential of the spin interaction can be analytically obtained and is dependent on the alignment of angular momenta of the black holes. For the collision of massive black holes, the gravitational radiation is numerically obtained as the upper bound by using the laws of thermodynamics. The effect of the Hayward black hole tends to increase the radiation energy, but we can limit the effect by comparing the radiation energy with the gravitational waves GW150914 and GW151226.}}

\thispagestyle{empty}
\newpage
\setcounter{page}{1}

\section{Introduction}

The collision of black holes is one of the ways by which black holes become larger in our universe. In particular, massive black holes, whose masses range from several to several tens of times the solar mass in recent detections at the Laser Interferometer Gravitational-Wave Observatory (LIGO)\cite{Abbott:2016blz,Abbott:2016nmj}, can drastically increase their masses through collision. For example, in the signal GW150914, two black holes having masses $36^{+5}_{-4} M_{\odot}$ and $29^{+4}_{-4} M_{\odot}$ merged to a single black hole of a mass $62^{+4}_{-4} M_{\odot}$, which is almost twice the mass of each of the initial black holes. In addition, during their collision, enormous energy was released in the form of gravitational radiation, and was detected at LIGO. By the serial detection of radiation from the collision, we can easily presume the existence of many massive black holes that had formed in the early stage of our universe and  grown their masses by colliding with one another.

These massive black holes might have played an important role as a gravitational impurity in the evolution of the early universe. Suggested by the discovery of the Higgs particle\cite{ATLAS:2012ae,Chatrchyan:2012tx}, one of possibilities is the metastability of the present universe in the studies of the Higgs potential\cite{Degrassi:2012ry,Blum:2015rpa}. Before decaying into the true vacua, the lifetime of this metastable stage can be so long as to be compatible with that of our universe, because a large energy barrier exists\cite{Coleman:1977py,Callan:1977pt,Coleman:1980aw}. Incidentally, the lifetime can be shortened to millions of Planck times by a gravitational impurity such as a black hole, which generates inhomogeneities that lower the energy barrier in the Higgs potential\cite{Burda:2015isa,Burda:2015yfa}. Therefore, an investigation into the collision of black holes is not only about the gravitational wave, but also about the early universe.

The black hole is also treated as a thermal system. The temperature of the black hole, called the Hawking temperature\cite{Hawking:1974sw,Hawking:1976de}, can be defined by its radiation through the quantum effect. In addition, a specific part of energies included in the mass of the black hole always increases despite the Penrose process\cite{Bardeen:1970zz,Penrose:1971uk}. The energy is called the irreducible mass of the black hole\cite{Christodoulou:1970wf,Christodoulou:1972kt,Smarr:1972kt}. The behavior of the irreducible mass is similar to that of the entropy in a thermal system, so that the entropy of the black hole can be obtained in the form of the square of the irreducible mass, which is proportional to the area of the black hole¡¯s horizon\cite{Bekenstein:1973ur,Bekenstein:1974ax}. Based on these thermal properties, thermodynamic laws are constructed for the black hole system. Applying thermodynamics, we can estimate the amount of gravitational radiation released during the collision of the black holes. Since the collision of black holes is an irreversible process, the entropy should increase from the initial to final states in the process, according to the second law of thermodynamics. For the case of the Schwarzchild black hole, the radiation energy is obtained in terms of the upper bound, which is sufficiently large to be included the observation detected nowadays at LIGO\cite{Hawking:1971tu}. For the case of rotating black holes, that is, Kerr black holes, the spin interaction between black holes contributes to the gravitational radiation, where dependencies on the alignments of the rotating axes exist between them\cite{Wald:1972sz}. If one of the black holes involved in the collision is small enough to be treated as a spinning particle, the upper bound of the radiation corresponds exactly to the potential of the spin interaction between black holes obtained from the Mathisson--Papapetrou--Dixon(MPD) equations for the spinning particle\cite{Schiff:1960gi,Mashhoon:1971nm,Wilkins:1970wap,Majar:2012fa,Zilhao:2013nda,Plyatsko:2015bia,d'Ambrosi:2015xci}. Hence, we can expect that the gravitational radiation  in the collision of the black holes will be affected by any interactions between black holes. In addition, the spin interaction may be changed in the higher-dimensional spacetime, and this change is also observable in the gravitational radiation released in the collision of higher-dimensional rotating black holes\cite{Myers:1986un,Herdeiro:2008en,Gwak:2016cbq}. Thus, we may estimate various aspects of black holes in gravity theories by investigating gravitational radiation, despite their upper bounds.

Studies in numerical relativity provide more precise estimation for gravitational radiation. The radiation released in a head-on collision of black holes was studied with the development of theoretical and computational frameworks in numerical relativity\cite{Smarr:1976qy,Smarr:1977fy,Smarr:1977uf,Witek:2010xi} based on the Einstein field equations. Then, various aspects of the gravitational radiation were studied and analyzed with different initial conditions\cite{Anninos:1993zj,Anninos:1998wt,Zilhao:2010sr,Witek:2014mha,Bednarek:2015dga,Hirotani:2015fxp,Sperhake:2015siy}. In particular, the waveforms of the gravitational waves, which were obtained from numerical-relativity simulation for the binary black hole merger, were produced in catalogs to be applied to estimate and study parameters related to events such as GW150914 and GW151226\cite{Mroue:2013xna,Ajith:2012az,Hinder:2013oqa,Husa:2015iqa,Lovelace:2016uwp,Abbott:2016wiq}. In addition, the detected gravitational waves were low-frequency waves, and they behaved in line with our rough expectations.

Further, in the near-horizon geometry of the black hole, the quantum effect becomes important owing to the strong gravity. One could expect that the geometry of the spacetime can be modified from the quantum effect  in the near-horizon region, so that the curvature singularity inside the black hole can be removed using the effect. Although various candidates and modifications considering quantum gravity exist still, the clear picture regarding quantum gravity is not well known. A regular black hole is one of the modifications focusing on removing a singularity, so that the black hole is regular in the whole spacetime. The Hayward black hole is one of regular black holes and a Bardeen-like black hole\cite{Bardeen:1968biy,Hayward:2005gi} given from the modified Einstein equations describing the formation and evaporation of the black hole from a vacuum. The modification is mainly applied to the near-horizon region of the black hole, and the extent to which the spacetime is modified from Einstein's spacetime is given as an additional parameter $g$ in the metric of the Hayward black hole, but far from the horizon, the black hole is approximately the same as the Schwarzschild black hole\cite{Hayward:2005gi,Frolov:2016pav}. Now, the rotating Hayward black hole is found by using the Newman--Janis transformation\cite{Bambi:2013ufa} in which the black hole corresponds to the Kerr black hole in the limit of $g$ going to zero or an asymptotic region. The horizon and the ergoregion of the rotating Hayward black hole are slightly different from those of the Kerr black hole in the dependency on $\theta$ direction\cite{Amir:2015pja,Amir:2016nti,DeLorenzo:2015taa}. The Hayward black hole is originally obtained in the modification of the Einstein equations mainly denoted as the parameter $g$, but the classes of regular black holes are found in gravity theories coupled with nonlinear electrodynamics, and the Hayward black hole is included in the classes\cite{AyonBeato:1998ub,AyonBeato:1999ec,AyonBeato:1999rg,AyonBeato:2000zs,Junior:2015fya,Fan:2016rih,Fan:2016hvf,Pourhassan:2016qoz}. In nonlinear electrodynamics, the source of regular black holes is the magnetic charge related to the parameter $g$ and different from the Schwarzschild black hole, which has a mass that gives rise to the singularity. Therefore, no singularity exists in a regular black hole in nonlinear electrodynamics.

In this paper, we have investigated the upper bounds on the gravitational radiation released in the collision of two rotating Hayward black holes. Regular black holes, including Hayward black holes, are candidates that can be described as astrophysical black holes, such as Cygnus X-1, using their deviation parameters\cite{Bambi:2011mj,Bambi:2014nta,Debnath:2015hea,Bambi:2017yoz}. However, the gravitational radiation released when regular black holes collide is not well studied despite serial detections at LIGO; hence, our results might be a reference for the further works on the gravitational wave from the models of regular black holes. In addition, the rotating Hayward black hole is an appropriate model to find out the extent to which the regularity effect in the near-horizon region of the black hole affects the radiation released from the black hole system, when we assume the regularity of the spacetime obtained from the modification of the Einstein equations. We have reduced the rotating Hayward black hole from three- to two-parameter systems and found the upper bound thermally allowed by using the numerical method owing to the $\theta$-dependency on the horizon of the black hole. Although the deviation of the Hayward black hole mainly affects the near-horizon geometry of the black hole, the effects can be observed in the upper bound on the gravitational radiation and spin interaction between two Hayward black holes. In addition, depending on the angle between angular momenta of the black holes in the collision, the upper bounds can be deviated from those of Kerr black holes due to the modification of the Hayward black hole given by the parameters. We have shown the extent to which the gravitational radiation depends on the parameters of the Hayward black hole and have found the effective range of the parameters by using the data from GW150914 and GW151226.

Note that we have mainly treated the parameter $g$ of the Hayward black hole as a universal constant in the spacetime, because the Hayward black hole is originally obtained from the modified Einstein equations, and the parameter is related only to an energy level in the near-horizon region of the black hole\cite{Hayward:2005gi}, and can work as a constant acting on the spacetime in this case. However, the Hayward black hole can also be a solution to the gravity theory coupled with nonlinear electrodynamics in which the parameter $g$ is not a universal constant, but a magnetic charge introduced in Appendix A.. Thus, each black hole has its own magnetic charge such as $g_1$ and $g_2$. The overall behaviors of the upper bounds are similar in both cases, but they also have differences in specific examples. This is a similar case in the instability for the Reissner--Nordstr{\"o}m-anti-de Sitter (RN--AdS) black hole depending on the gravity model. In Einstein--Maxwell gravity, the RN-AdS black hole is stable under perturbation, but, in $\mathcal{N}=8$ gauged supergravity, the RN-AdS black hole is unstable under the scalar mode perturbation\cite{Konoplya:2008rq,Gubser:2000ec,Gubser:2000mm}. Hence, the radiation might be dependent on the gravity models for the Hayward black holes. For the cases of the nonlinear electrodynamics, we have provided a review and reproduced our results about the upper bounds on the radiations in Appendix A. 

The paper is organized as follows. In section~\ref{sec2}, we review rotating Hayward black holes. In section~\ref{sec3}, we focus on the contribution of the spin interaction in the upper bound on the gravitational radiation when one of the black holes is considerably smaller than the other. In section~\ref{sec4}, we describe our framework for the bound and numerically investigate the radiation bounds on the collision of massive black holes under the effect of deviation parameters given as specific ranges by using two LIGO data. In section~\ref{sec5}, we briefly summarize our results.

\section{Rotating Hayward Black Holes}\label{sec2}
\begin{figure}[h]
\centering\subfigure[{The $\theta$-directional dependency of the outer horizon at $g=0.3$, $M=1$, and $a=0.68$\,.}]
{\includegraphics[scale=0.65,keepaspectratio]{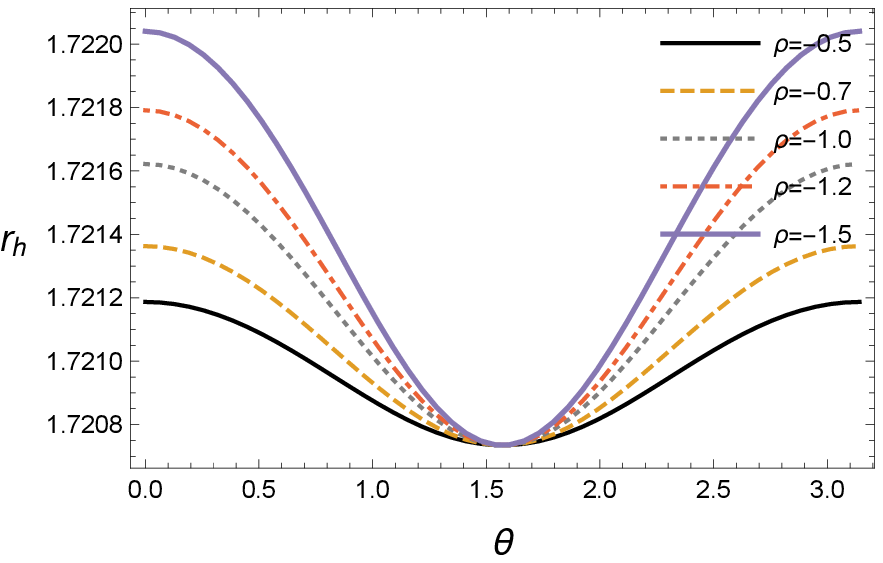}}
\quad
\centering\subfigure[{The outer horizon with respect to $g$ at $\rho=-1$, $M=1$, and $a=0.68$\,.}]
{\includegraphics[scale=0.65,keepaspectratio]{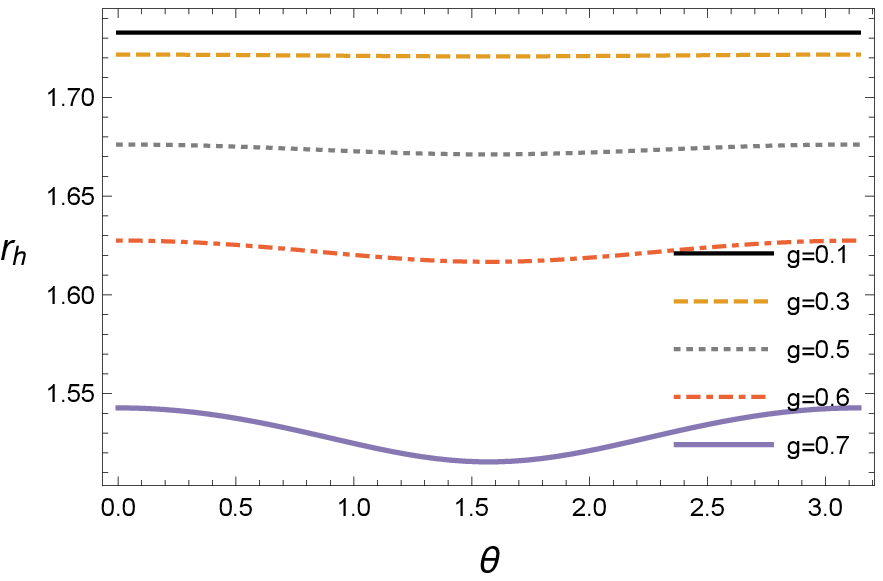}}
\quad
\centering\subfigure[{The surface under $\rho=-1$ and $g=0.3$, $M=1$, and $a=0.68$\,.}]
{\includegraphics[scale=0.27,keepaspectratio]{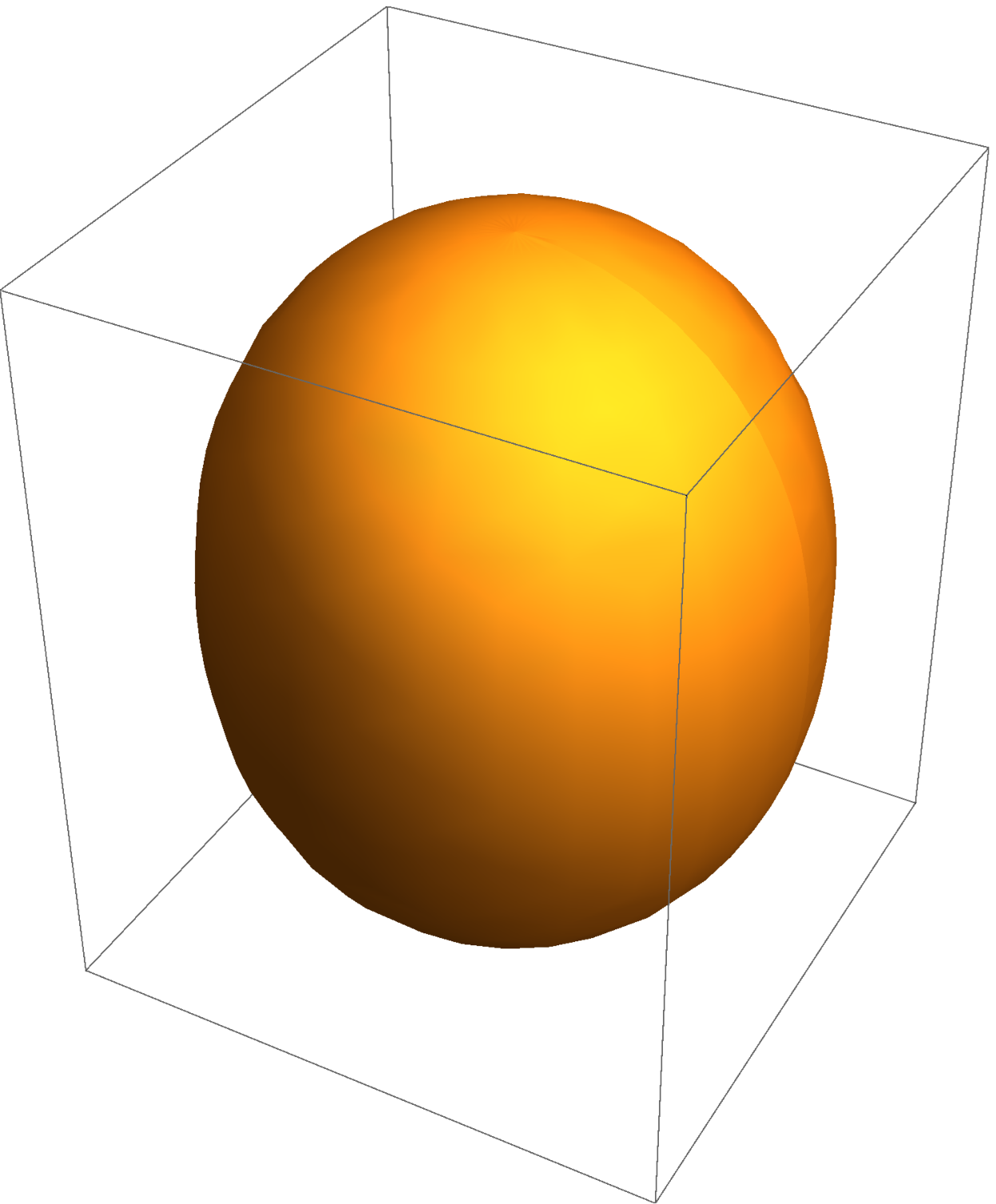}}
\caption{{\small The outer horizon and its surface for a negative $\rho$\,.}}
\label{fig:f3}
\end{figure}

The rotating Hayward black hole is obtained using the Newman--Janis transformation\cite{Bambi:2013ufa} from the Hayward black hole obtained from the modified Einstein equations\cite{Hayward:2005gi}. The rotating Hayward black hole is also a regular black hole that has no curvature singularity in the whole spacetime, as given by the Boyer--Lindquist coordinates
\begin{align}\label{eq:metric}
ds^2 &= - \left(1-\frac{2mr}{\Sigma}\right)dt^2- \frac{4 a m r\sin^2\theta}{\Sigma}dt d\phi + \frac{\Sigma}{\Delta}dr^2+\Sigma d\theta^2 +\left(r^2+a^2+\frac{2a^2 m r\sin^2\theta}{\Sigma}\right)\sin^2 \theta d\phi^2\,,\\
\Delta &= r^2-2 m r +a^2\,,\quad \Sigma = r^2+a^2\cos^2\theta\,,\nonumber
\end{align}
where the mass function is
\begin{eqnarray}
m =M \frac{r^{3+\alpha}\Sigma^{-\alpha/2}}{r^{3+\alpha} \Sigma^{-\alpha/2}+g^3 r^\beta \Sigma^{-\beta/2}}\,.\nonumber
\end{eqnarray}
The mass of the black hole is given as $M$, and the spin parameter is $a$. The angular momentum of the black hole is defined as $J=Ma$. The metric of the black hole is modified in the near-horizon region from the regularity effect. The extent to which the black hole is constructed by modifying the Einstein equations is denoted by the real deviation parameters $\alpha$, $\beta$, and $g$. The parameter $g$ describes the deviation of the energy level in the near-horizon region and is defined as positive. For simplicity, we reduce two parameters, $\alpha$ and $\beta$, to one parameter $\rho$ defined as $\rho=\alpha-\beta$ in Eq.~(\ref{eq:metric}). Then, the mass function is written as,
\begin{eqnarray}\label{eq:massform}
m=M \frac{r^{3+\rho}}{r^{3+\rho}+g^3\Sigma^{\rho/2}}\,.
\end{eqnarray}
Therefore, we will use parameters $\rho$ and $g$ without loss of generality in this work. The rotating Hayward black hole is regular in the whole spacetime\cite{Bambi:2013ufa}. The rotating Hayward black hole is recovered to the Kerr black hole at $g=0$, where there is no effect of the modified Einstein equations working in the near-horizon region of the black hole. In addition, the black hole becomes a Schwarzschild black hole at $g=0$ and $a=0$\cite{Amir:2015pja,Amir:2016nti,DeLorenzo:2015taa}. The horizon of the rotating Hayward black hole can be obtained from $\Delta$ in Eq.~(\ref{eq:metric}) that depends on the radial and $\theta$ coordinates, so that the horizon depends on not only the coordinate $r$, but also the coordinate $\theta$, except $\rho=0$\,.
\begin{figure}[h]
\centering\subfigure[{The $\theta$-directional dependency of the outer horizon at $g=0.3$, $M=1$, and $a=0.68$.}]
{\includegraphics[scale=0.65,keepaspectratio]{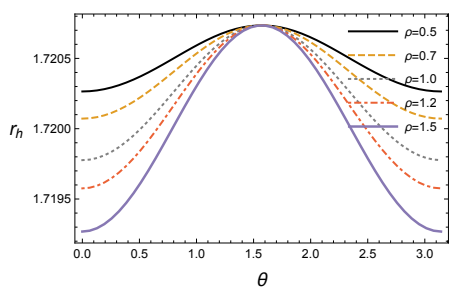}}
\quad
\centering\subfigure[{The outer horizon with respect to $g$ at $\rho=1$, $M=1$, and $a=0.68$\,.}]
{\includegraphics[scale=0.65,keepaspectratio]{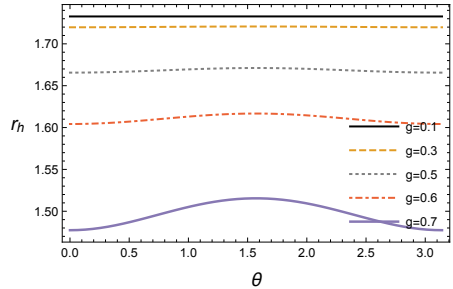}}
\quad
\centering\subfigure[{The surface of the outer horizon under $\rho=1$, $g=0.3$, $M=1$, and $a=0.68$\,.}]
{\includegraphics[scale=0.3,keepaspectratio]{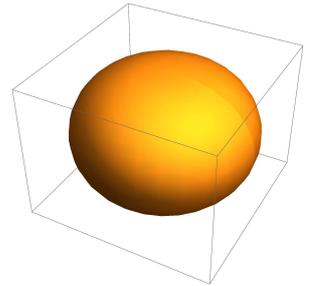}}
\caption{{\small The outer horizon and its surface for a positive $\rho$\,.}}
\label{fig:f1a}
\end{figure}
This is different from the Kerr black hole having a sphere-like surface of the horizon. The properties of the outer horizon can be classified into two cases along with the sign of $\rho$, as shown in Fig.~\ref{fig:f1a} and \ref{fig:f3}. Their surfaces and extremal condition are changed according to the sign of $\rho$\,.

\begin{figure}[h]
\centering\subfigure[{The extremal spin parameter with respect to $\theta$ at $g=0.3$ and $M=1$\,.}]
{\includegraphics[scale=0.6,keepaspectratio]{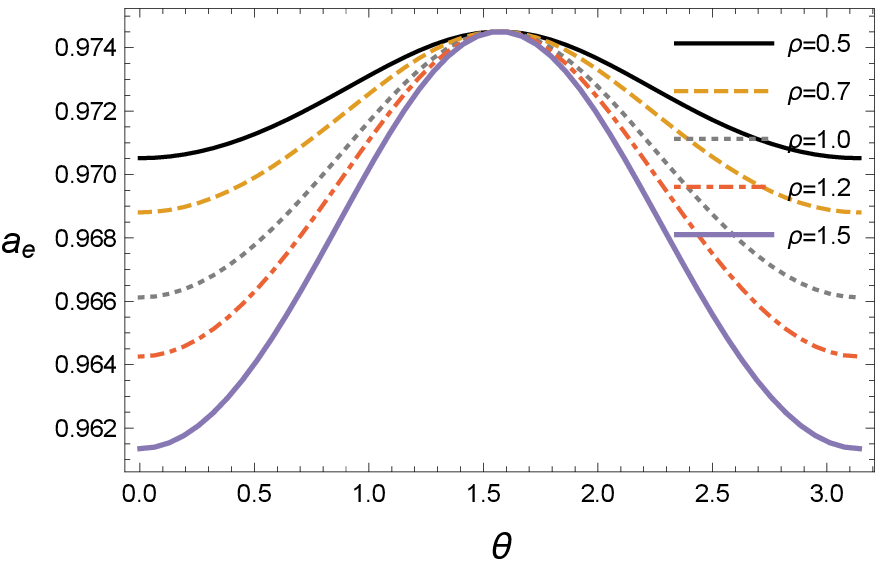}}
\quad
\centering\subfigure[{The extremal spin parameter with respect to $\theta$ at $\rho=1$ and $M=1$\,.}]
{\includegraphics[scale=0.6,keepaspectratio]{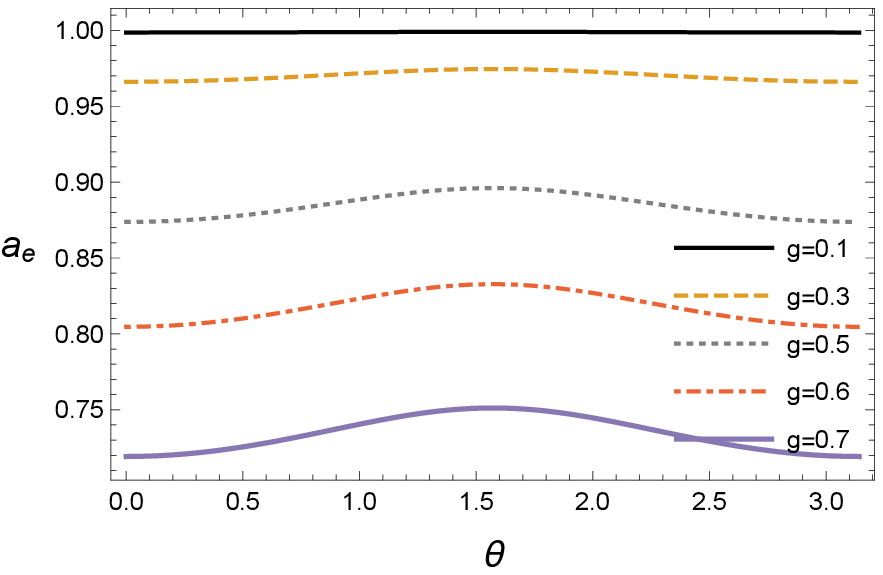}}
\quad
\centering\subfigure[{The minimum extremal spin parameter with respect to $\rho$ and $g$ under $\rho=1$ and $M=1$\,.}]
{\includegraphics[scale=0.6,keepaspectratio]{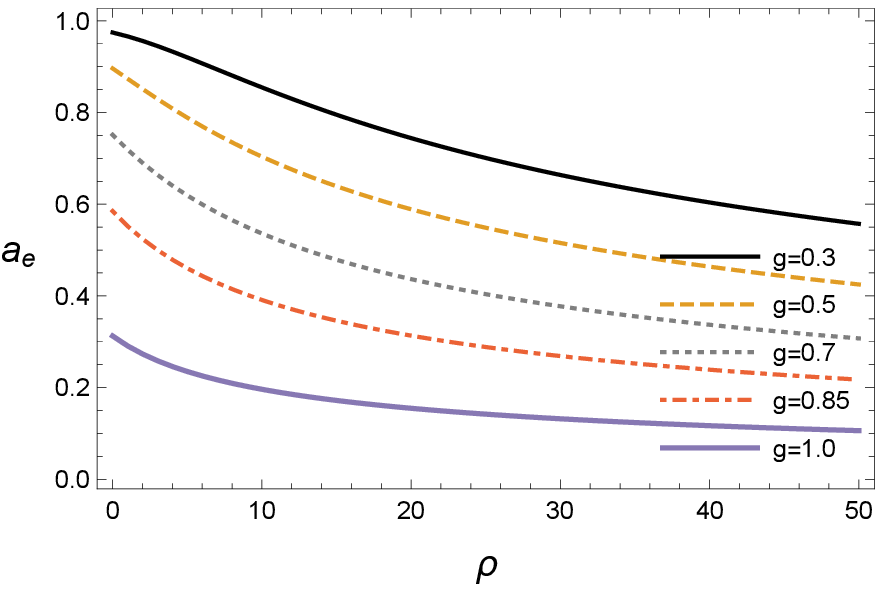}}
\caption{\small The extremal spin parameter for a positive $\rho$.}
\label{fig:f2b}
\end{figure}

For a positive value of the parameter $\rho$, the location of the outer horizon depends on the $\theta$ direction, and the surface of the horizon is not a sphere in this coordinate system. Thus, it is different from that of the Kerr black holes, as shown in Fig.~\ref{fig:f1a}~(a). The horizon also depends on the parameter $g$ and becomes small at large values of $g$ due to the large regularity effects in Fig.~\ref{fig:f1a}~(b). The outer horizon is the minimum at $\theta=0,\,\pi$ and the maximum at $\theta=\pi/2$. The difference between parameters is related to their effects on the black hole. The parameter $\rho$ changes the locations of the horizon at $\theta=0,\,\pi$, but the parameter $g$ changes the whole locations of the horizon. The surface of the horizon is expected as Fig.~\ref{fig:f1a}~(c). For the positive cases, the north and south poles are shorter than the equator. Similar to the location of the horizon, the extremal spin parameter $a_e$ is also different for a given $\theta$ direction. At the extremal spin parameter, the inner and outer horizons are coincident to each other, and the temperature of the black hole becomes zero. For the given parameters $g$ and $\rho$, the extremal spin parameter is the minimum at $\theta=0,\,\pi$ and the maximum at $\theta=\pi/2$, as shown in Fig.~\ref{fig:f2b}. The parameter $\rho$ changes the extremal spin parameter except for $\theta=\pi/2$ in Fig.~\ref{fig:f2b}~(a), and the extremal spin parameter reacts sensitively to the change of the parameter $g$. In the larger parameter $g$, the extremal spin parameter becomes smaller in Fig.~\ref{fig:f2b}~(b).
Since the extremal spin parameter is different in the $\theta$ coordinate for the given $g$ and $\rho$ parameters, in this work, we define the extremal spin parameter for the positive $\rho$ case as the value at the north and south poles, the minimum value, because the Hayward black hole can be still a black hole below the minimum extremal spin parameter. The extremal spin parameter depends on values of $\rho$ and $g$ as shown in Fig.~\ref{fig:f2b}~(c). As the value of $\rho$ increases, the extremal spin parameter decreases, and hence the range of the spin parameters becomes tight.

\begin{figure}[h]
\centering\subfigure[{The extremal spin parameter with respect to $\theta$ at $g=0.3$ and $M=1$\,.}]
{\includegraphics[scale=0.6,keepaspectratio]{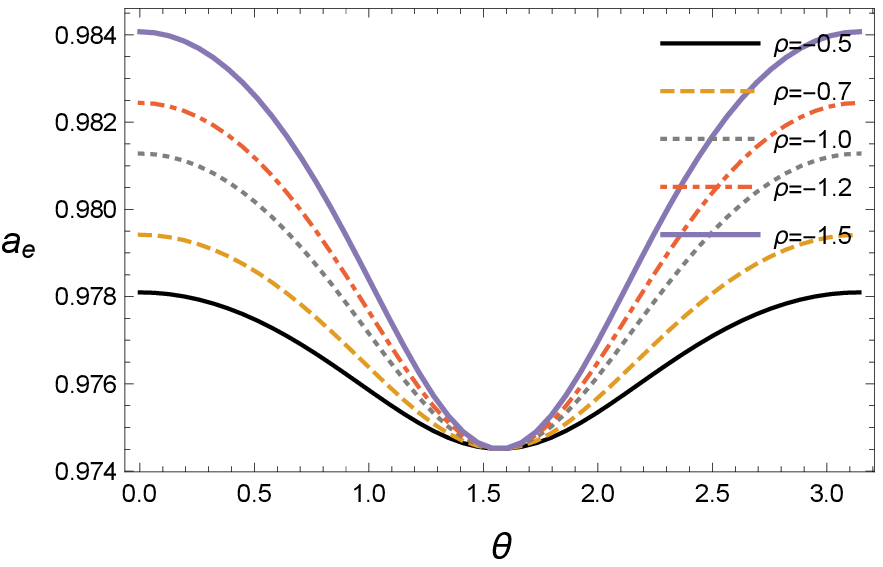}}
\quad
\centering\subfigure[{The extremal spin parameter with respect to $\theta$ at $\rho=-1$ and $M=1$\,.}]
{\includegraphics[scale=0.6,keepaspectratio]{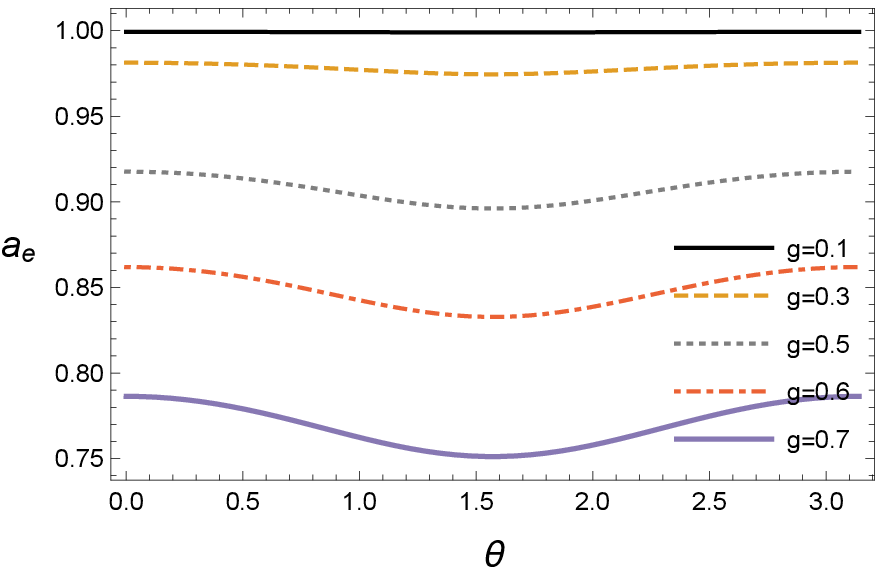}}
\quad
\centering\subfigure[{The minimum extremal spin parameter for given $\rho$ and $g$ at $M=1$\,.}]
{\includegraphics[scale=0.6,keepaspectratio]{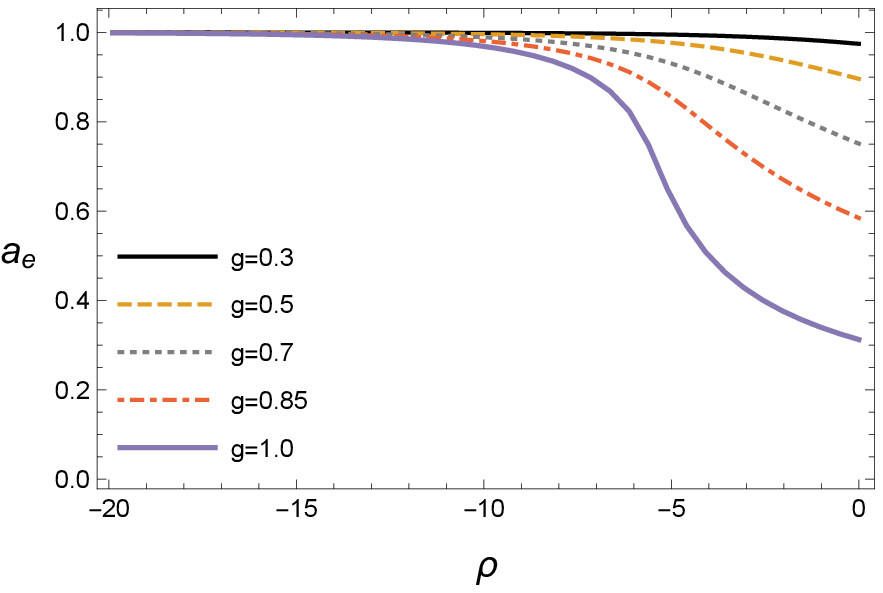}}
\caption{{\small The extremal spin parameter for a negative $\rho$\,.}}
\label{fig:f4ae}
\end{figure}

For the negative sign of the parameter $\rho$, all properties are opposite to the positive cases, as shown in Fig.~\ref{fig:f3}. The outer horizon is the maximum at $\theta=0,\,\pi$ and the minimum at $\theta=\pi/2$ in Fig.~\ref{fig:f3}~(a). In addition, a change of the parameter $\rho$ varies the location of the horizon, except for that of $\theta=\pi/2$. The locations of the horizon are also sensitive to the change of the parameter $g$. The horizon becomes small at large values of the parameter $g$, as shown in Fig.~\ref{fig:f3}~(b). The horizon is long at the north and south poles of the black hole and short at the equator, and the surface of the horizon looks as shown in Fig.~\ref{fig:f3}~(c), which is the opposite of the positive $\rho$ case.
The extremal spin parameter $a_e$ is also opposite to that of the positive cases in $\theta$-directional behaviors, as shown in Fig.~\ref{fig:f4ae}. At the north and south poles of the black hole, the extremal spin parameters are the maximum, and at the equator, the extremal spin parameter is the minimum, as shown in Fig.~\ref{fig:f4ae}~(a). For the change of the parameter $g$, the extremal spin parameter also becomes small as seen in Fig.~\ref{fig:f4ae}~(b).
The definition of the extremal spin parameter for the negative $\rho$ case is also similar to that for the positive case. For the given $\rho$ and $g$, we set the extremal spin parameter at the value of the equator. Further, the spin parameter of the black hole should be below the minimum extremal spin parameter to still be considered a black hole. The extremal spin parameter is still dependent on the value of $\rho$ and $g$ for the negative case as shown in Fig.~\ref{fig:f4ae}~(c), where the extremal spin parameter decreases as in the positive cases. However, for larger $g$, the change is drastic with respect to the change in $\rho$. If the spin parameter is larger than its extremal value, the horizon disappears by parts beyond the value. This is no longer a black hole, and hence we will not be concerned with collisions between these kinds of objects. Future work in which their thermal property is investigated might be interesting. However, for completeness, we have briefly tested the objects, as reported below.

\begin{figure}[h]
\centering\subfigure[{The outer horizons for $\rho=1$, $g=0.3$, and $M=1$\,.}]
{\includegraphics[scale=0.6,keepaspectratio]{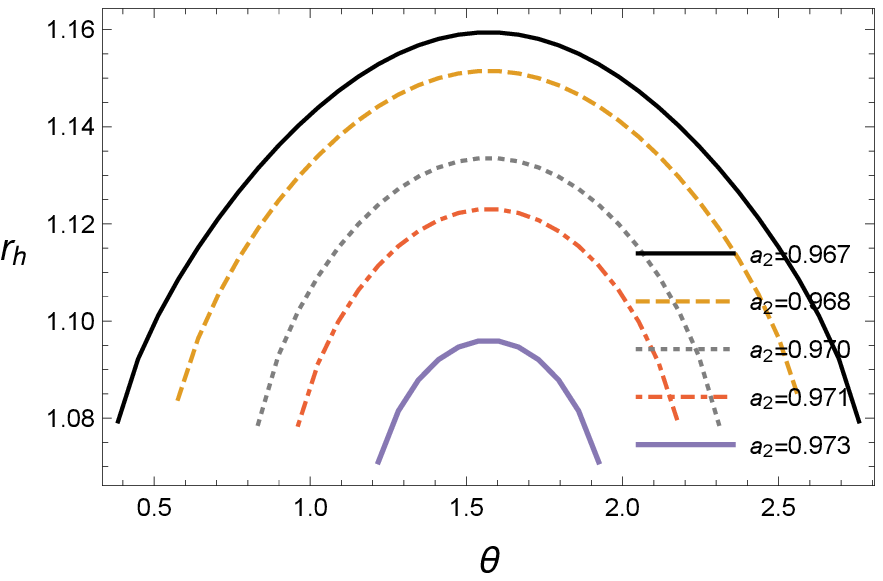}}
\quad
\centering\subfigure[{The function $\Delta$ in the $\theta$ direction for $\rho=1$, $g=0.3$, and $M=1$\,.}]
{\includegraphics[scale=0.6,keepaspectratio]{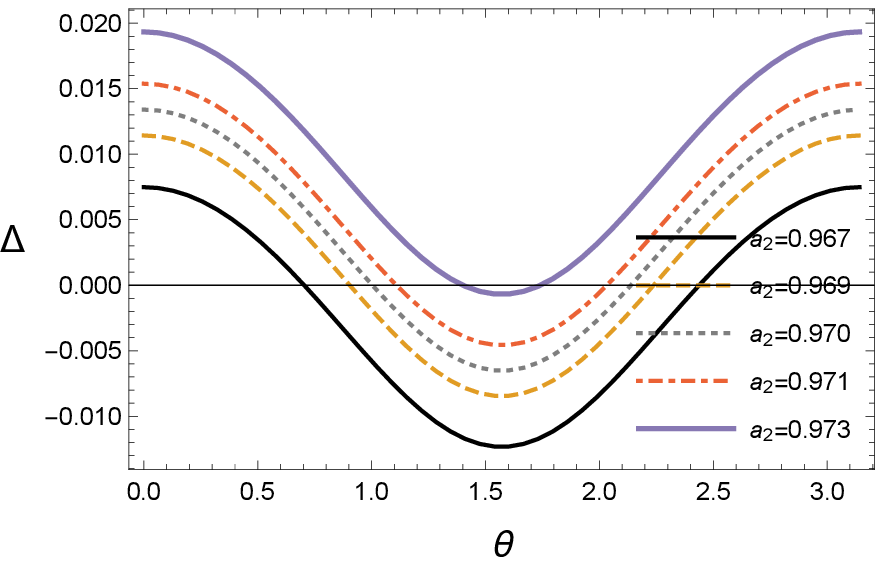}}
\quad
\centering\subfigure[{The surface of the outer horizon for for $\rho=1$, $g=0.3$, $M=1$, and $a=0.97$\,.}]
{\includegraphics[scale=0.3,keepaspectratio]{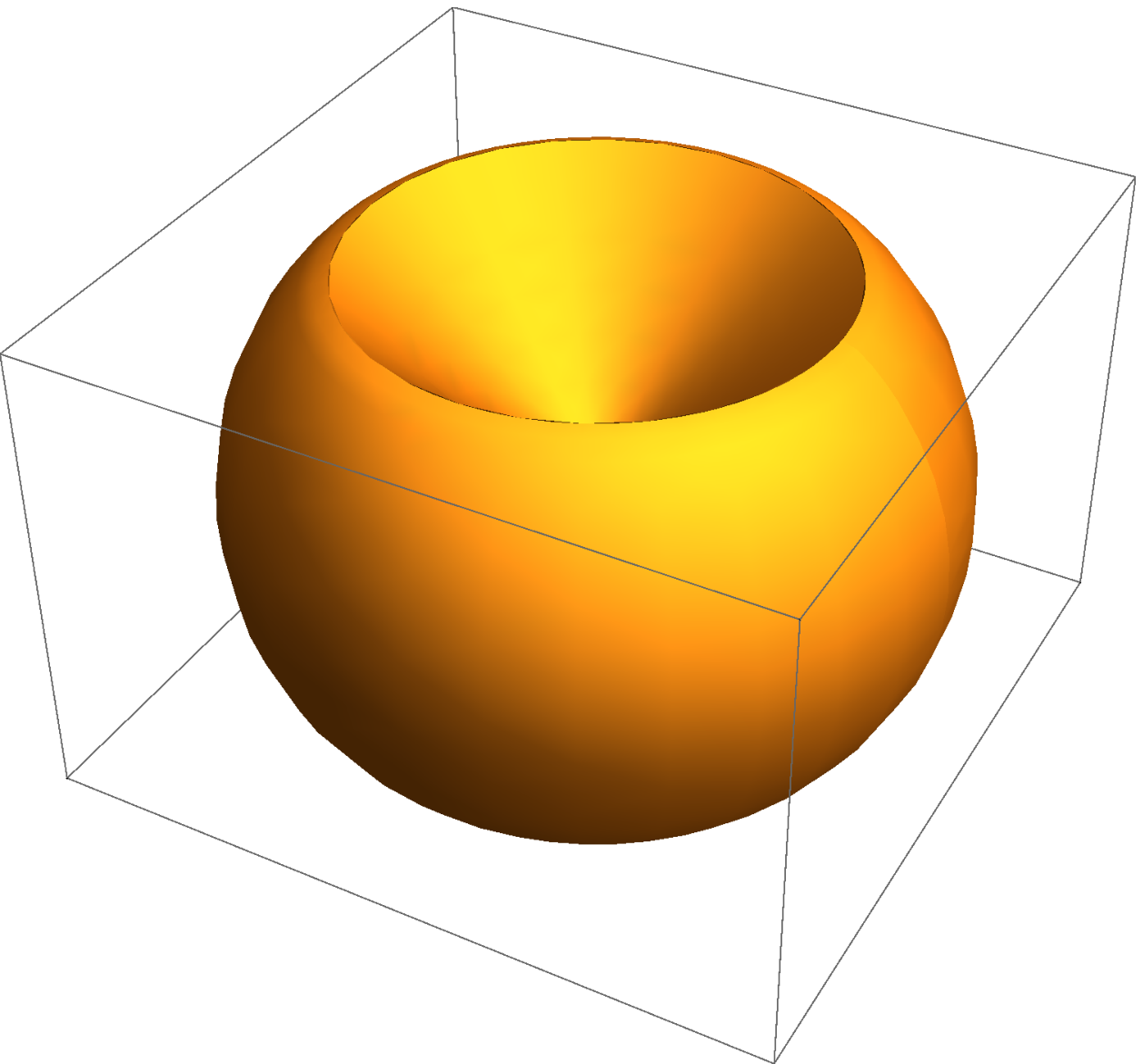}}
\caption{{\small The outer horizons of overspinning cases for a positive $\rho$\,.}}
\label{fig:f5a}
\end{figure}

If the spin parameter exceeds the extremal value $a_e$, the horizon starts disappearing at the points having a smaller extremal value, not at the same time, as shown in Fig.~\ref{fig:f5a}~(a) for a positive $\rho$. Then, the inside of the black hole is observable through the open area. However, this situation does not violate the weak cosmic censorship that prevents a naked singularity in our universe\cite{Penrose:1969pc,Wald:1974ge}, because there is no singularity in its inside.
The extremal spin parameter is the smallest at the pole of the black hole for positive $\rho$ cases, so that the horizon first disappears at the pole in the radial direction. However, there are horizon-like surfaces at $\theta$ directions in Fig.~\ref{fig:f5a}~(b) because the function of $\Delta$ still has two sign changes in the $\theta$ direction. This means that even if the horizon surface opens to the radial direction, we cannot observe the back of the horizon. Considering the $\theta$-directional surface, the surface of the horizon can be as that shown in Fig.~\ref{fig:f5a}~(c) to the asymptotic observer.

\begin{figure}[h]
\centering\subfigure[{The outer horizons for $\rho=-1$, $g=0.3$, and $M=1$\,.}]
{\includegraphics[scale=0.6,keepaspectratio]{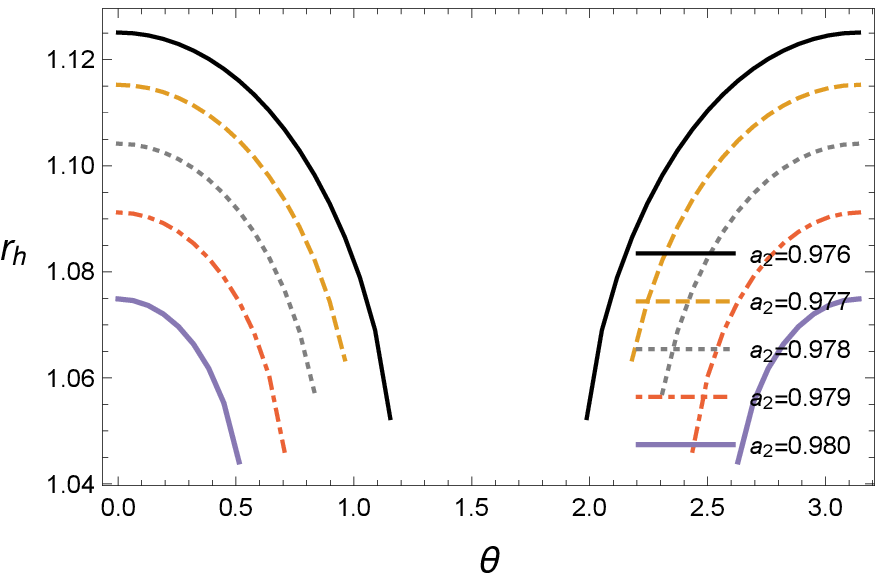}}
\quad
\centering\subfigure[{The function $\Delta$ in the $\theta$ direction for $\rho=-1$, $g=0.3$, and $M=1$\,.}]
{\includegraphics[scale=0.6,keepaspectratio]{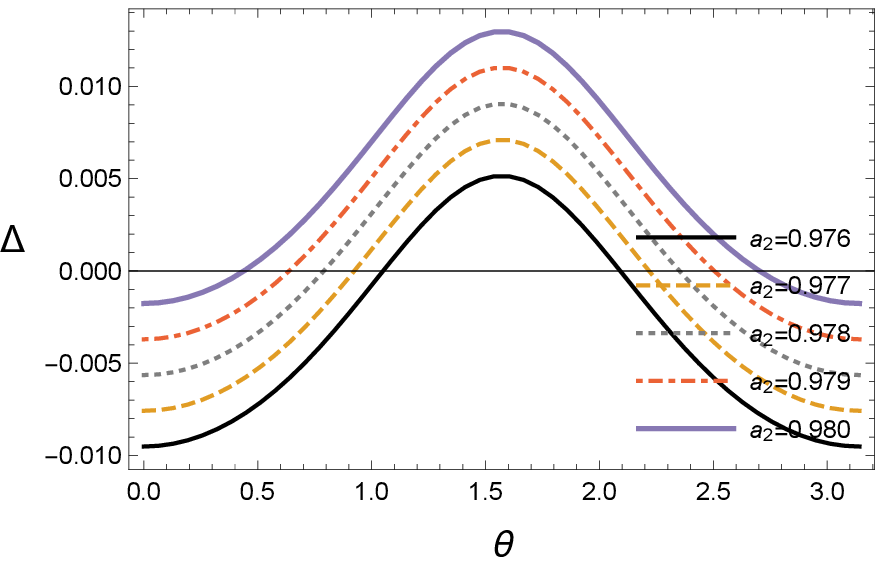}}
\quad
\centering\subfigure[{The horizon surface for $\rho=-1$, $g=0.3$, $M=1$, and $a=0.97$\,.}]
{\includegraphics[scale=0.3,keepaspectratio]{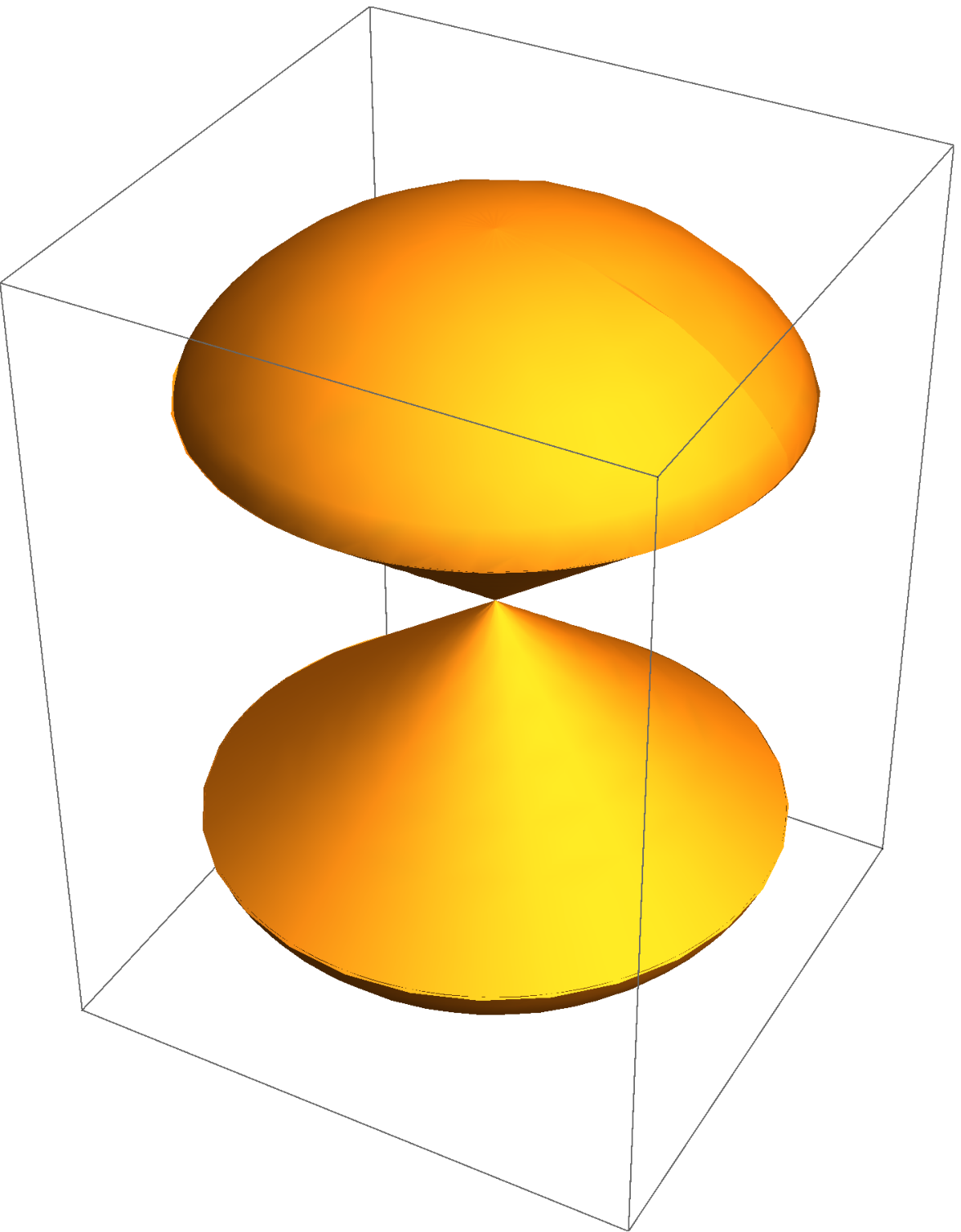}}
\caption{{\small The outer horizons of overspinning cases for a negative $\rho$\,.}}
\label{fig:f6a}
\end{figure}

For the negative $\rho$ cases, their appearance is opposite to the positive ones, as shown in Fig.~\ref{fig:f6a}~(a). Their extremal spin parameter is the smallest at the equator, so that the horizon starts to disappear at the equator.
Hence, the coverage of the horizon on the $\theta$ direction becomes small along with the increase in the spin parameter. In addition, as we have seen in the case of positive $\rho$, even if there is no horizon in the radial direction, the $\theta$-directional horizon exists and forms a closed surface instead of the radial direction, as shown in Fig.~\ref{fig:f6a}~(b), so that the observer cannot see the whole spacetime of the black object. Then, the region veiled by the horizon will look as it does in Fig~\ref{fig:f6a}~(c).

\begin{figure}[h]
\centering\subfigure[{The area for $\rho=1$ and $g=0.3$\,.}]
{\includegraphics[scale=0.6,keepaspectratio]{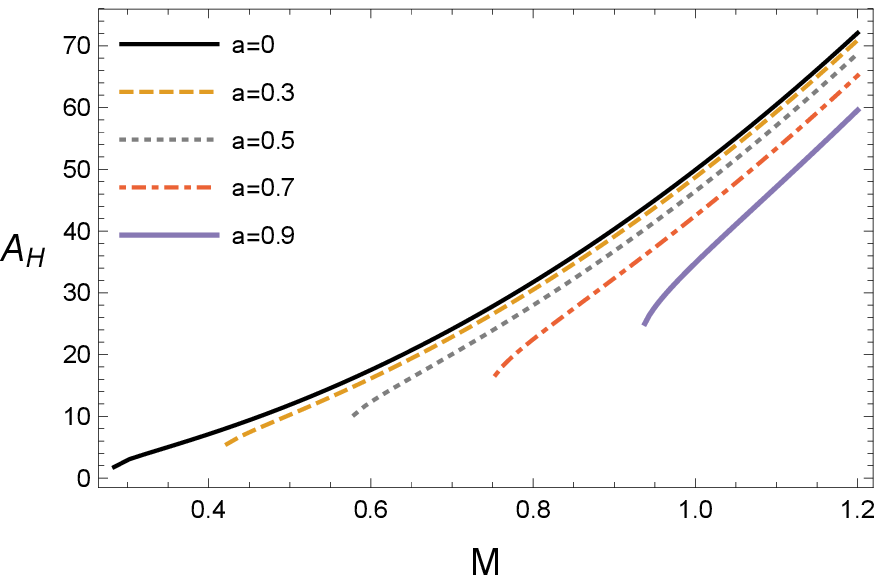}}
\quad
\centering\subfigure[{The area with respect to $\rho$ for $g=1$\,.}]
{\includegraphics[scale=0.6,keepaspectratio]{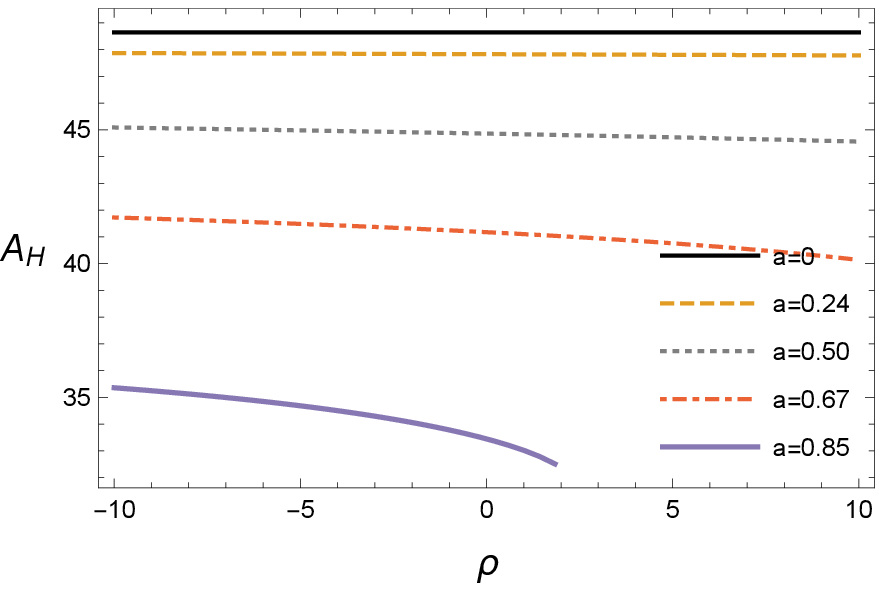}}
\quad
\centering\subfigure[{The area with respect to $g$ for $\rho=1$\,.}]
{\includegraphics[scale=0.6,keepaspectratio]{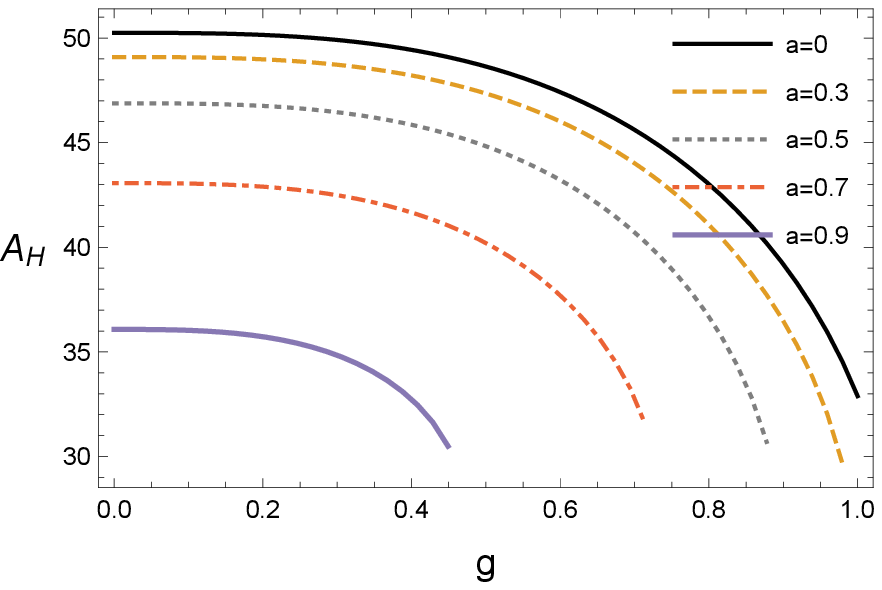}}
\caption{{\small The area of the outer horizon of the rotating Hayward black hole.}}
\label{fig:f7b}
\end{figure}

In this work, we mainly focus on the upper bounds of the thermally allowed gravitational radiation in the collision of two Hayward black holes. The procedure of this work will be based on the second law of thermodynamics. This will be described in section~\ref{sec4}. The entropy of the black hole is given by the Bekenstein--Hawking entropy $S_{BH}$, which is proportional to the horizon area $A_H$ of a black hole, so that
\begin{eqnarray}
S_{BH}=\frac{1}{4}A_H\,,\quad A_H=2\pi\int_{0}^{\pi} \left(r_h^2(\theta)+a^2\right) \sin\theta d \theta\,,
\end{eqnarray}
where the area of the horizon surface is obtained from the metric component $g_{\phi\phi}$ and $g_{\theta\theta}$ at the horizon, a function of $\theta$. However, we do not have an exact form of the horizon at a given $\theta$; hence, the area of the horizon will be numerically obtained in this work. The overall behaviors of the area are shown in Fig.~\ref{fig:f7b}.
The area of the Hayward black hole is similar in the response of changes in mass and spin parameter, as shown in Fig.~\ref{fig:f7b}~(a). As the value of $\rho$ increases, the area decreases in Fig.~\ref{fig:f7b}~(b), but zero-spin parameter does not depend on $\rho$, because the effect of $\rho$ is removed in $m$ in these cases. The effects of $g$ are greater than those of $\rho$, as Fig.~\ref{fig:f7b}~(c) shows. For the given parameters, the area of the black hole becomes small as the parameter $g$ increases. Since the Kerr black hole is at $g=0$, the Hayward black hole can be differentiated at large values of $g$, so that the differences between Hayward and Kerr black holes can be obtained at large values of $g$.

\section{Spin Interaction in Rotating Hayward Black Hole}~\label{sec3}
Spin interaction acts between two objects having a spin angular momentum. Two rotating black holes are also coupled with each other by a spin interaction through which the potential energy can be released through the gravitational radiation. Before investigating into massive Hayward black holes, we will obtain the potential of the spin interaction between a Hayward black hole and a spinning particle, and then the potential will correspond to the energy of the gravitational radiation in the collision of two Hayward black holes when one of the black holes is small enough compared with the other. We suppose that a spinning particle has a spin interaction coupled with the angular momentum $J_1$ of a rotating black hole having mass $M_1$ and spin parameter $a_1$. The interaction acts as attractive or repulsive to the particle, and can be shown using the equations of motion described by the MPD equations\cite{Mathisson:1937zz,Papapetrou:1951pa,Dixon:1970zza}. For the spinning particle having mass $m_2$ and four-velocity $v^\mu$, the MPD equations are given as in the black hole spacetime
\begin{eqnarray}
\frac{D p^a}{D s} = -\frac{1}{2}R^a_{bcd}v^bS^{cd}\,,\quad \frac{D S^{ab}}{D s} = p^a v^b-p^b v^a\,,
\end{eqnarray} 
where the Riemann curvature tensor of the black hole is $R^a_{bcd}$\,. The spin tensor of the particle $S^{ab}$ is defined as the difference between its momentum $p^a$ and velocity $v^a$ with respect to the proper time $s$\. The spin tensor is related to the spin vector of the particle $S_a$ as
\begin{eqnarray}
S_a=\frac{1}{2m_2}\sqrt{-g}\epsilon_{abcd}\,p^bS^{cd}\,.
\end{eqnarray}
To obtain the trajectory of the spinning particle, we chose a supplementary condition\cite{Beiglbock1967}, and then, the magnitude of the spin $J_2$ and mass $m_2$ of the particle are obtained as
\begin{eqnarray}\label{eq:suppl}
p_a S^{ab}=0\,,\quad J_2^2=\frac{1}{2}S_{ab} S^{ab}\,,\quad \frac{D S^{ab}}{D s} =0\,,\quad m_2^2=-p_ap^a\,,\quad p^a = m_2 v^a\,.
\end{eqnarray}
We simply assume an initial condition where the spinning particle slowly comes into the pole of the rotating Hayward black hole, and their rotating planes are parallel to each other. The normalized velocity $v^a$ and spin vector $S^a$ are given as follows under the initial condition:
\begin{eqnarray}
v^{a}=\left(\frac{1}{\sqrt{-g_{tt}}},\,v^a,\,0,\,0\right)\,,\quad J_2^a=\left(0,\,\frac{J_2}{\sqrt{g_{rr}}},\,0,\,0\right)\,,
\end{eqnarray}
where the slowly moving particle can be treated as a nonrelativistic particle; hence $v^r\ll1$\,. In the initial condition, the energy of the spinning particle $E$ is obtained as a conserved quantity for the Killing vector of the time direction $\xi^t$. Then,
\begin{eqnarray}
E=-p_t-\frac{1}{2}S^{ab}\nabla_a g_{bt}\,,
\end{eqnarray}
where the first term is the energy of the nonspinning particle, and the effect of the spinning particle comes from the potential of the spin interaction $U_{spin}$ given in the second term. Due to the pole-to-pole collision, the potential is obtained at the pole of the black hole $\theta=0$ and horizon $r=r_1$:
\begin{eqnarray}\label{eq:potential01}
U_{spin}=\frac{2r_1^{\rho+4}J_1 J_2 }{\left(r_1^2+a_1^2\right)^2\left(r_1^{\rho+3}+g^3\left(r_1^2+a_1^2\right)^{\rho/2}\right)}\,,
\end{eqnarray}
where the angular momentum of the rotating black hole is denoted as $J_1=M_1a_1$\,. The potential of the spin interaction in Eq.~(\ref{eq:potential01}) shows that the spinning particle undergoes attraction for $J_1 J_2<0$ and repulsion for $J_1 J_2 >0$\,. The anti-parallel alignment of the rotating planes between the particle and the black hole has a negative potential acting as attraction, and their parallel alignments have a positive potential acting as repulsion. Therefore, if we assume that a part of the interaction energy is released as gravitational radiation, the anti-parallel alignment may radiate more energy than the parallel alignment, because the anti-parallel alignment has a negative potential.

The potential of the Hayward black hole in Eq.~(\ref{eq:potential01}) corresponds to that of the Kerr black hole in the limit of $g=0$. Then, in the limit, the potential is given as
\begin{eqnarray}\label{eq:potential05}
U_{spin}=\frac{2r_1J_1 J_2 }{\left(r_1^2+a_1^2\right)^2}\,,
\end{eqnarray}
which is that of the Kerr black hole\cite{Wald:1972sz,Gwak:2016cbq}. Therefore, our result is consistent with the spin interaction obtained in the solution of the Einstein gravity.

\section{Gravitational Radiation under Collision of Hayward Black Holes}~\label{sec4}
The energy of the spin interaction certainly contributes to the gravitational radiation released in the collision of two rotating black holes. When the mass of one black hole is negligible in comparison with the other, most of the radiation energy can be released from the spin interaction energy, which is a reducible energy in the black hole system. In massive black holes, the effect of the mass becomes important, and the radiation energy originated from the mass will be greater than that from the interaction. We now introduce the general procedure to obtain the upper bound on the gravitational radiation generalized from the case of the Schwarzschild black hole\cite{Hawking:1971tu}, and then we will estimate the approximate potential of the spin interaction and the maximum of the gravitational radiation in the cases of small mass of a black hole and arbitrary alignments between black holes. In addition, we will apply the upper bound for the collision between massive black holes and discuss the features of Hayward black holes.

We assume an initial state when two Hayward black holes stay far from each other. The first black hole has mass $M_1$ and angular momentum $J_1$. $M_2$ and $J_2$ are the mass and angular momentum of the second black hole, respectively. The gravitational interaction is ignored in the initial state, because the black holes are separated by a long distance. Then, attracted to each other, the black holes collide head-on and form a Hayward black hole having mass and angular momentum $M_3$ and $J_3$ in the final state. If the final black hole has a smaller mass than the sum of masses of the initial black holes, the loss of mass in the final state can be expected to be released through the gravitational radiation. Hence, the energy of the gravitational radiation $M_{r}$ is obtained from
\begin{eqnarray}\label{eq:radiation01}
M_r=(M_1+M_2)-M_3\,.
\end{eqnarray}
The collision of black holes is an irreversible process; therefore, the entropy should increase in the final state with respect to that in the initial state, according to the second law of thermodynamics. We suppose that the area of the outer horizon is denoted as $A_H(M_1,J_1)$ for the first black hole, $A_H(M_2,J_2)$ for the second one, and $A_H(M_3,J_3)$ for the final black hole. Since the Bekenstein--Hawking entropy is proportional to the area of the horizon, the areas should satisfy
\begin{eqnarray}\label{eq:inequality01}
A_H(M_1,J_1)+A_H(M_2,J_2)\leq A_H(M_3,J_3)\,,
\end{eqnarray}
where the entropy of the gravitational radiation is not included in that of the final state, because the radiation is a very small portion of the mass for the initial state\cite{Hawking:1971tu,Wald:1972sz}. For example, the released energy is expected to be approximately 4\% of the total mass in GW150914 and GW151226\cite{Abbott:2016blz,Abbott:2016nmj}. Hence, the consideration for the entropy of the radiation may not change the upper bound for the radiation energy. In addition, we will show that the results are reasonable without the consideration for the entropy of the radiation. In particular, in the limit of the small mass, the bound on the radiation will give exact results. The mass of the final black hole can be obtained by solving Eq.~(\ref{eq:inequality01}) and imposing the conservation of the angular momentum in the final state. Then,
\begin{eqnarray}\label{eq:angular}
\vec{J}_1+\vec{J}_2=\vec{J}_3\,,
\end{eqnarray}
because the radiation may be released in an arbitrary direction. Then, net angular momentum can be conserved in this process, so that the angular momentum of the final black hole is still the sum of the angular momenta in the initial state without any loss. In the final state, the magnitude of the angular momentum will be one of variables that determine the minimum of the horizon at the extremal condition. When the angle between angular momenta $\vec{J}_1$ and$\vec{J}_2$ is defined as $\psi$, the vector sum of the angular momenta is
\begin{eqnarray}\label{eq:angular05}
J_3 = \sqrt{J_1^2 + J_2^2 + 2J_1J_2 \cos\psi}\,,
\end{eqnarray}
where the parallel alignment is for $\psi=0$, and the anti-parallel alignment is for $\psi=\pi$\,. From the inequality in Eq.~(\ref{eq:inequality01}), the minimum mass of the final black hole $M_{3,min}$ is obtained in terms of given initial states. Then, from Eq.~(\ref{eq:radiation01}), the upper bound on the radiation $M_{r,bound}$ is the thermally allowed maximum energy released for a given initial state. Therefore, the real radiation should be inside the upper bound
\begin{eqnarray}\label{eq:inequalityrad01}
M_{r,bound} \geq M_r\,.
\end{eqnarray}
Most of the energy beyond the inequality in Eq.~(\ref{eq:inequalityrad01}) occurs owing to the radiation energy contributed from the mass. In the inequality in Eq.~(\ref{eq:inequalityrad01}), the effect of the black hole mass may be too great to be considered in $M_{r,bound}$. Hence, the upper bound is greater than that of the observation in the real collisions detected at LIGO. Therefore, if we assume one of black holes in the initial state to be infinitesimally small, the effect of the mass can be removed in $M_{r,bound}$, so that the extent to which the spin interaction affects the radiation can be clearly described in the collision of Hayward black holes.

\subsection{Analytical Approximation in Limit of Small Mass}
The collision of black holes can release the gravitational radiation in which various interaction energies of the black hole system are included. The energy of the spin interaction is one of these energies included in the spinning black hole system. In the collision, if one black hole is considerably smaller than the other, the radiation from the interaction can be important in that, following the mass loss of the initial black holes, the upper bound of the radiation in Eq.~(\ref{eq:inequalityrad01}) will be close to the exact value. In the limit of the small mass, we assume that the second black hole is sufficiently small in comparison with the first black hole, and hence $M_1\gg M_2$\,. In addition, the angular momentum of the second black hole should be small in the mass scale, and hence $M_1^2\gg J_2$\,. To compare the gravitational radiation with the potential of the spin interaction in Eqs.~(\ref{eq:potential01}) and (\ref{eq:inequalityrad01}), we consider the pole-to-pole collision of the black holes after calculating an arbitrary angle $\psi$. Under the initial condition, the final black hole mass will be obtained to satisfy the second law of thermodynamics in Eq.~(\ref{eq:inequality01}), so that the area of the final black hole should be greater than the sum of the areas of the initial black holes. However, the area of the Hayward black hole is obtained by a numerical integration owing to the dependency on the $\theta$ coordinate and written as a numerical function. To find an analytical equation, the radius of the outer horizon is assumed to be a fixed value at $\theta=0$ to remove the dependency on the $\theta$ direction, because the radius of the horizon varies for the $\theta$ direction, as shown in Figs.~\ref{fig:f1a} and \ref{fig:f3} . Then, the first black hole is supposed to be the larger one having the radius of the outer horizon $r_1$ and the second black hole has the radius $r_2$, and $r_1\gg r_2$\,. After the collision, the final black hole will have the horizon radius $r_3$ satisfying Eq.~(\ref{eq:inequality01}) rewritten as,
\begin{eqnarray}\label{eq:modifiedeq1}
4\pi (r_1^2+a_1^2)+4\pi (r_2^2+a_2^2) \leq 4\pi (r_3^2+a_3^2)\,,
\end{eqnarray}
where we write down the area in a form similar to that of the Kerr black hole for consistency with the $g=0$ case\cite{Wald:1972sz}. The minimum mass of the final black hole or the upper bound of the gravitational radiation is obtained from Eq.~(\ref{eq:radiation01}), by imposing the conservation in Eq.~(\ref{eq:angular}) and the equality of Eq.~(\ref{eq:modifiedeq1})\,. In addition, each value of the horizon radius is the solution of the function $\Delta$ at the pole of $\theta=0$\,
\begin{eqnarray}
\Delta {\big{|} }_{\theta=0}=0\,.
\end{eqnarray}
Because the angular momentum of the second black hole $J_2$ is small enough, the upper bound on the gravitational radiation can be obtained in terms of the partial derivative with respect to $J_2$\,. Our derivation is valid in the limit of $g$ going to zero, because we assume the form of the area is as the case of the Kerr black hole in Eq.~(\ref{eq:modifiedeq1}). Therefore,
\begin{eqnarray}\label{eq:potential03}
\frac{\partial M_{r,bound}}{\partial J_2}=-\frac{|J_1| \cos\psi}{M_1 \left(r_1^2+\left(\frac{J_1}{M_1}\right)^2\right)}-\frac{|J_1| \left(r_1^2+\left(\frac{J_1}{M_1}\right)^2\right)^{\rho/2} (3 + \rho) \cos\psi}{M_1 r_1^{ 1 + \rho} \left(r_1^2+\left(\frac{J_1}{M_1}\right)^2\right)^2}g^3+\mathcal{O}(g^6)\,,
\end{eqnarray}
in which we consider the magnitudes of $J_1$ and $J_2$, because their angle difference is given in $\psi$. The spin interaction is a force with respect to the angular momenta of the two black holes instead of the displacement, so that the change in the radiation energy can be related to the interaction force. The negative sign in Eq.~(\ref{eq:potential03}) indicates that the interaction between black holes is attractive for the anti-parallel alignment and repulsive for the parallel alignment. We can obtain the interaction potential by treating Eq.~(\ref{eq:potential03}) as a conserved force with respect to $J_2$\,. Then, the negative sign is removed. The potential of the spin interaction is obtained from Eq.~(\ref{eq:potential03}) under the limit of $g=0$ as
\begin{eqnarray}\label{eq:kerrlimit05}
U_{spin}=\frac{2 |J_1| |J_2| r_1 \cos\psi}{(r_1^2+a_1^2)^2}+\frac{2|J_1||J_2|(3+\rho)\cos\psi}{r_1^{\rho}(r_1^2+a_1^2)^{3-\rho/2}}g^3+\mathcal{O}(g^6)\,,
\end{eqnarray}
in which the first term imposed to $\psi=0$ is that of the Kerr black hole in Eq.~(\ref{eq:potential05})\cite{Wald:1972sz,Gwak:2016cbq}, and the second term concerns the effect of the Hayward black hole. Hence, we can estimate that the potential energy corresponds to the upper bound of the gravitational radiation in the collision. However, the second term is different from the expansion of Eq.~(\ref{eq:potential01}) because the increase of the area in Eq.~(\ref{eq:modifiedeq1}) is not an exact equation and mimics the Kerr black hole. Therefore, gravitational radiation is released as much as the potential of the spin interaction for which the amount of energy is approximately that of the Kerr black hole in the limit of the small mass. Thus, a more exact effect of the Hayward black hole in the radiation should be obtained numerically under the consideration of its exact area rather than Eq.~(\ref{eq:kerrlimit05}).

\subsection{Numerical Analysis in Massive Cases}
\begin{figure}[h]
\centering\subfigure[{The upper bounds for $\psi$ at the different $M_2$ under $\rho=1$, $g=0.5$, $M_1=1$, $a_1=0.35$, and $a_2=0.1$\,.}]
{\includegraphics[scale=0.9,keepaspectratio]{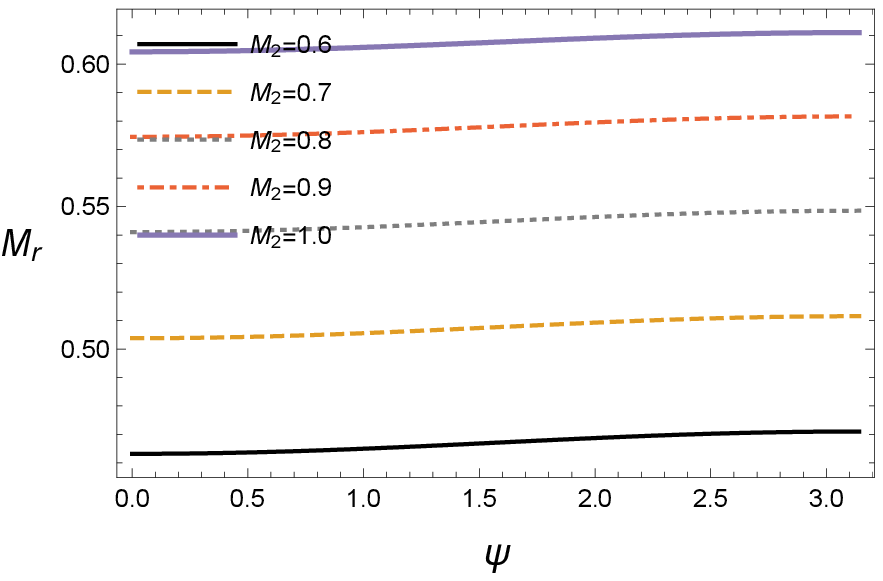}}
\quad
\centering\subfigure[{The upper bounds for $\psi$ at the different $a_2$ under $\rho=1$, $g=0.5$, $M_1=1$, $a_1=0.35$, and $M_2=1$\,.}]
{\includegraphics[scale=0.9,keepaspectratio]{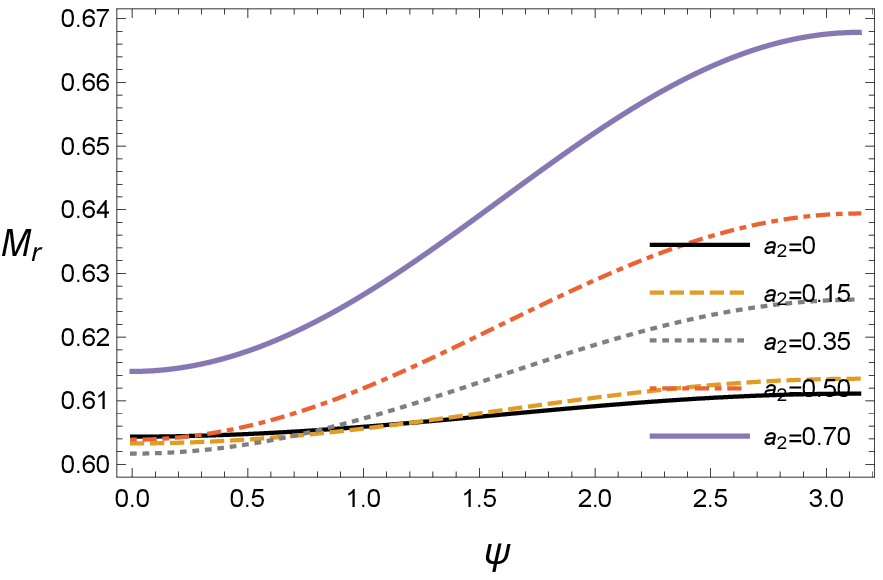}}
\caption{{\small The upper bound on the gravitational radiation for a positive $\rho$\,.}}
\label{fig:f8}
\end{figure}

In the collision of massive black holes, the gravitational radiation and their interaction should be investigated numerically, because we cannot assume one of the black holes as a particle and ignore the effect of its mass. The upper bound on the radiation is obtained from Eq.~(\ref{eq:radiation01}) using the minimum mass of the final black hole in Eq.~(\ref{eq:inequality01}) by imposing the conservation of the angular momentum in Eq.~(\ref{eq:angular}). In this work, we assume that the behavior of the real radiation $M_r$ is proportional to the upper bound on the radiation $M_{r,bound}$, so that we will use $M_r$ instead of $M_{r,bound}$ for simplicity. In addition, to apply the numerical method, we scale all parameters having mass dimension by the mass of the  most massive black hole rather than the other, so that mass of the most massive black hole is set to unity and $M_1$.
\begin{figure}[h]
\centering\subfigure[{The upper bounds for $M_2$ at $\rho=1$, $g=0.5$, $M_1=1$, $a_1=0.35$, and $\psi=0$\,.}]
{\includegraphics[scale=0.9,keepaspectratio]{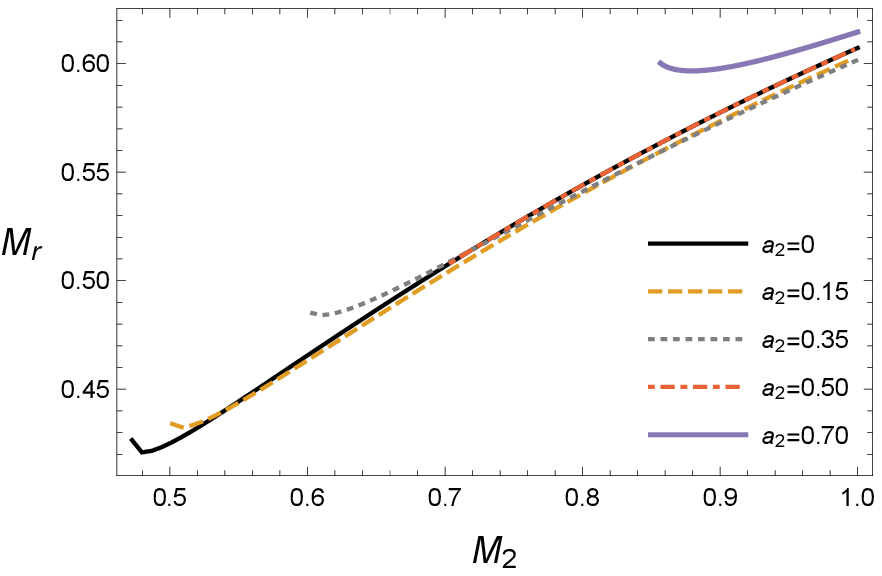}}
\quad
\centering\subfigure[{The upper bounds for $a_2$ at $\rho=1$, $g=0.5$, $M_1=1$, $M_2=1$, and $\psi=0$\,.}]
{\includegraphics[scale=0.9,keepaspectratio]{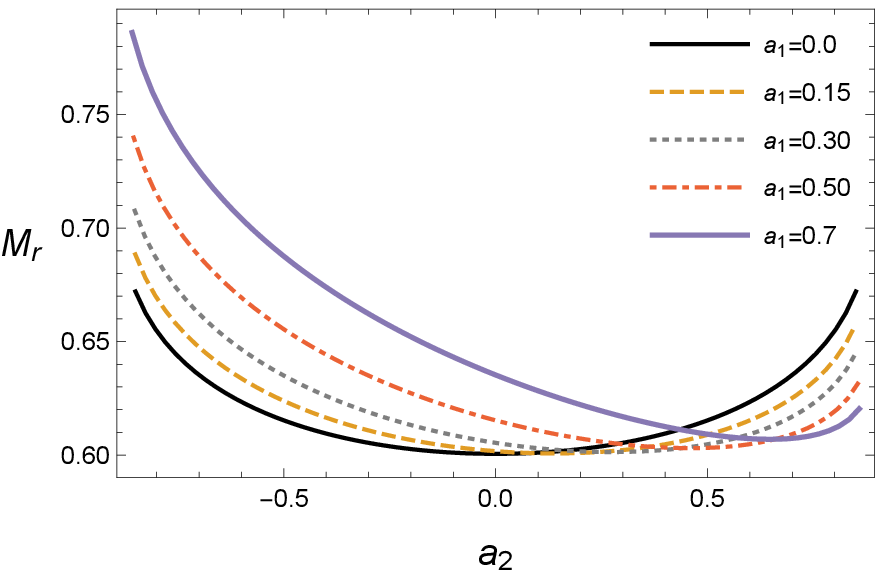}}
\caption{{\small The upper bound on the gravitational radiation for a positive $\rho$\,.}}
\label{fig:f7}
\end{figure}
In the mass scaling, the mass of the second black hole will not exceed unity of the mass of the first black hole, because we can set the mass of the most massive black hole to unity by the scaling. Various parameters are related to the Hayward black hole, and we first show the upper bound on the radiation with respect to parameters such as $M_1$, $a_1$, $M_2$, $a_2$, and $\psi$, which determine the initial state for fixed parameters $g$ and $\rho$. Then, we will show dependencies on parameters $g$ and $\rho$. The upper bounds on the gravitational radiation are obtained in Fig.~\ref{fig:f7} with respect to the second black hole $M_2$ and $a_2$ for the given first black hole $M_1$ and $a_1$ in $g\,,\rho>0$ cases. The radiation $M_r$ increases as the mass of the black hole increases, as shown in Fig.~\ref{fig:f7}~(a). For a given spin parameter $a_2$\,, the radiation starts at the minimum mass of $M_2$, because the Hayward black hole has an extremal bound related to $g$ and $a$. In addition, the minimum mass of $M_2$ for the $a_2=0$ case is also related to the extremal bound for $g$, which is different from the case for the Kerr black hole. The radiation also depends on parameters $a_1$ and $a_2$ for fixed $M_1$ and $M_2$ in Fig.~\ref{fig:f7}~(b), where the angle $\psi$ is fixed to $0$, so that the positive spin parameter is in parallel alignment, and the negative spin parameter is in anti-parallel alignment. For the same magnitude for spin parameter $a_2$\,, the anti-parallel case releases more radiation than the parallel case, because the slope, $\frac{\partial M_{r,bound}}{\partial J_2}$, is negative with respect to $J_2$. Since the source of the radiation is the energy of the black hole system, greater energy of the initial state can be extracted and released by the radiation due to the negative potential of the anti-parallel alignment in Eq.~(\ref{eq:potential01}) rather than the positive potential from Eq.~(\ref{eq:potential01}), even if the system has the same energy in the initial condition. Therefore, the maximum radiation occurs at the extremal spin parameter $a_e$ in the anti-parallel case. For the same reason, the minimum radiation is released at the parallel alignment having a positive potential. The angle $\psi \not= 0$ between the angular momenta of the initial black holes affects the radiation, as shown in Fig.~\ref{fig:f8}.
The magnitude of the angular momentum of the final black hole is given as the vector sum in Eq.~(\ref{eq:angular05}), and the angular momentum of the final black hole is not parallel to that of the initial black holes. Owing to the potential of the spin interaction, the radiation is minimum at the angle $\psi=0$ for the parallel alignment, and it becomes greater when the angle $\psi$ goes to $\psi=\pi$ for the anti-parallel alignment. The mass of the black hole and radiation energy increase together, as shown in Fig.~\ref{fig:f8}~(a). In addition, the radiation grows bigger as the spin parameter $a_2$ increases, as shown in Fig.~\ref{fig:f8}~(b), which is consistent with Fig.~\ref{fig:f7} ~(b), so that the anti-parallel cases release more energy than the parallel cases because of the contribution of the spin interaction.
\begin{figure}[h]
\centering\subfigure[{The upper bounds for $a_2$ at the different $a_1$ under $\rho=-1$, $g=0.5$, $M_1=1$, and $M_2=1$\,.}]
{\includegraphics[scale=0.9,keepaspectratio]{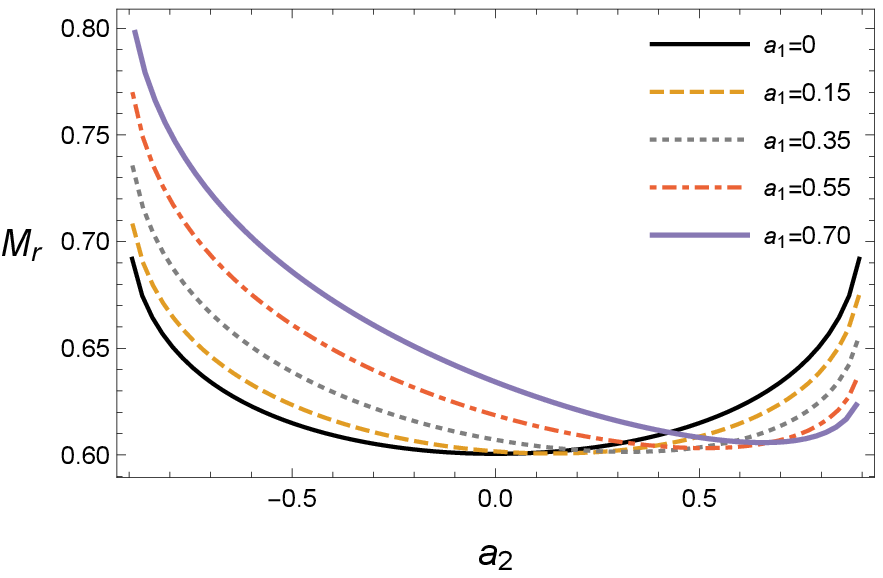}}
\quad
\centering\subfigure[{The upper bounds for $\psi$ at the different $a_2$ under $\rho=-1$, $g=0.5$, $M_1=1$, and $M_2=1$\,.}]
{\includegraphics[scale=0.9,keepaspectratio]{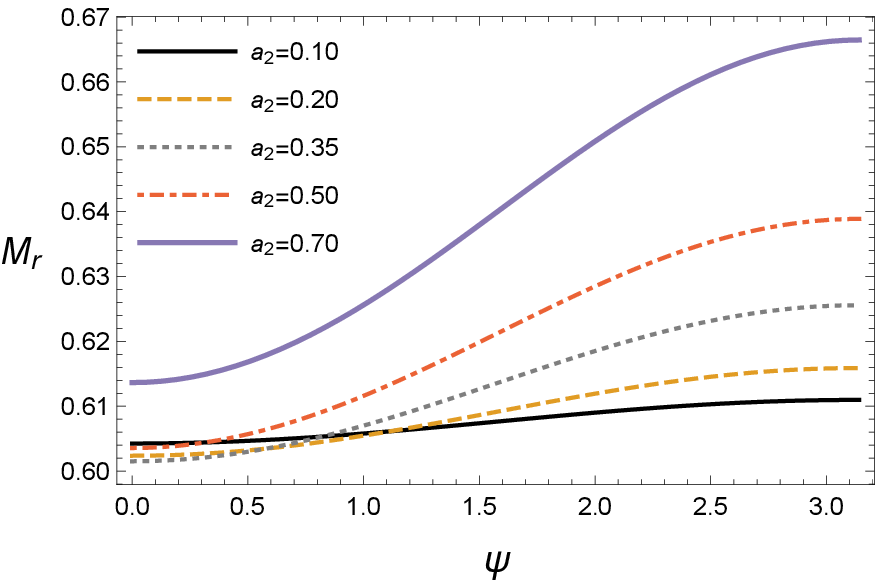}}
\caption{{\small The upper bound on the gravitational radiation for a negative $\rho$\,.}}
\label{fig:f7n}
\end{figure}
Gravitational radiation is also dependent on parameters $g$ and $\rho$, which are important constants to form a regular black hole. The parameters are fixed for a given spacetime. The radiation also depends on the parameters, but the dependency on the parameter $g$ is greater than that of $\rho$. For example, for the same value of $g$ in Fig.~\ref{fig:f7} and \ref{fig:f8}, the change in the radiation for negative $\rho$ is too small to ascertain, as shown in Fig.~\ref{fig:f7n}, where the radiation for a negative $\rho$ is slightly smaller than that of the positive $\rho$.

\begin{figure}[h]
\centering\subfigure[{{The upper bounds for $\rho$ at the different $M_2$ under $g=0.5$, $M_1=1$, $a_1=0.35$, and $a_2=0.5$\,.}}]
{\includegraphics[scale=0.9,keepaspectratio]{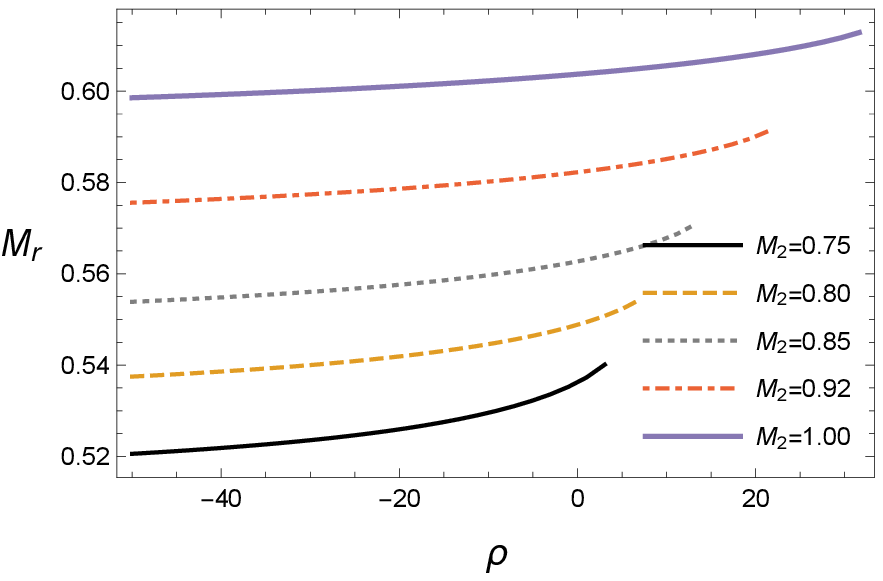}}
\quad
\centering\subfigure[{{The upper bounds for $\rho$ at the different $a_2$ under $g=0.5$, $M_1=1$, $a_1=0.35$, and $M_2=1$\,.}}]
{\includegraphics[scale=0.9,keepaspectratio]{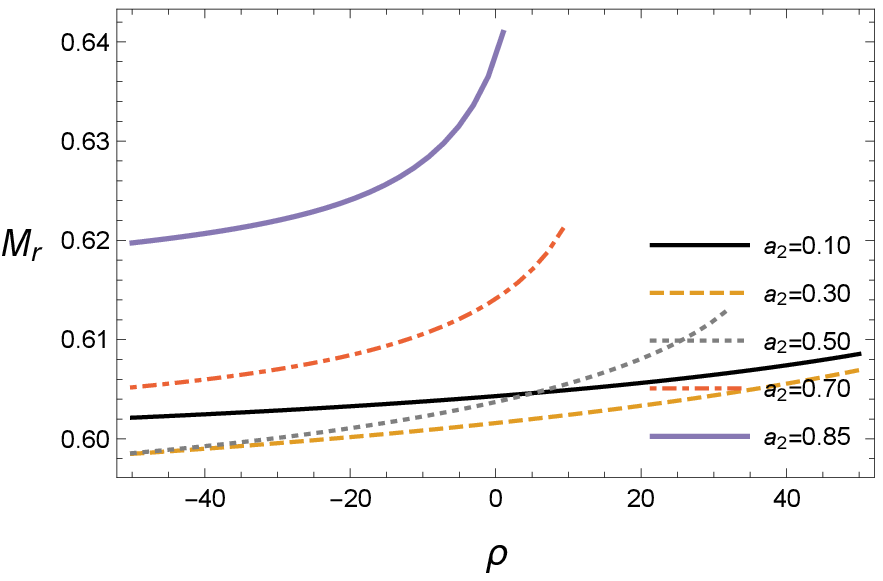}}
\caption{{\small The upper bound on the gravitational radiation with respect to $\rho$\,.}}
\label{fig:f91}
\end{figure}

The parameter $\rho$ only works in the mass function $m_\rho(r,\theta)$, so that the change in the power of $r$ by $\rho$ affects a very small portion of the mass function. However, the extremal condition for the spin parameter $a$ depends on the parameter $\rho$, and hence the negative and positive values of $\rho$ are distinguishable in the radiation, as shown in Fig.~\ref{fig:f91}.
The radiation increases as the value of $\rho$ becomes large, but the extremal bound on the spin parameter becomes smaller at large values of $\rho$. Then, at a large enough value of $\rho$, no initial black hole satisfying the extremal condition exists. Therefore, the large-mass black hole covers larger values of $\rho$ than the small mass black hole, as seen in Fig.~\ref{fig:f91}~(a). In addition, for a given spin parameter, the radiation also increases as the value of $\rho$ increases. However, the extremal bound on the spin parameter becomes smaller at large value of $\rho$; hence, at a large enough value of $\rho$, the radiation is limited, and no initial condition satisfying the extremal bound exists.
\begin{figure}[h]
\centering\subfigure[{{The upper bounds on the radiation for $g=0.7$, $M_1=1$, $a_1=0.15$, and $M_2=1$\,.}}]
{\includegraphics[scale=0.9,keepaspectratio]{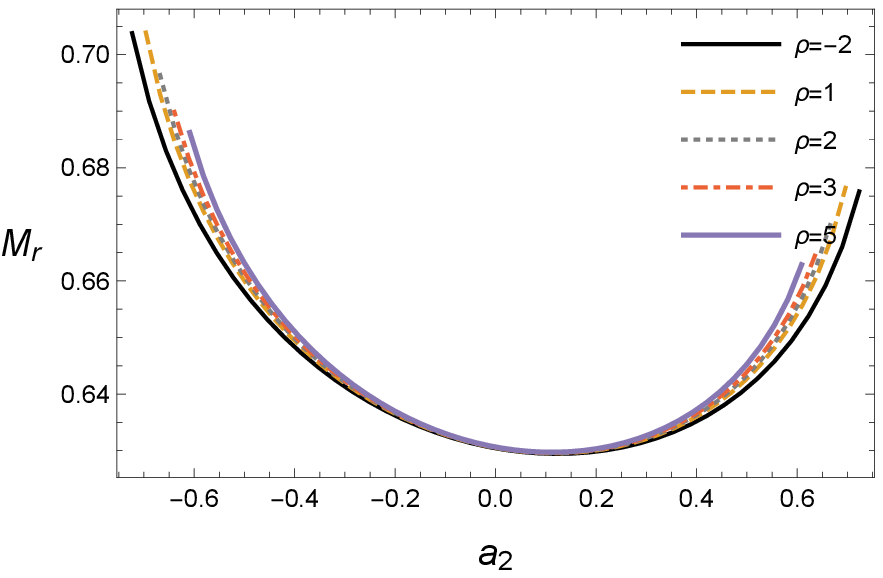}}
\quad
\centering\subfigure[{{The upper bounds on the radiation for $g=0.7$, $M_1=1$, $a_1=0.35$, and $a_2=0$\,.}}]
{\includegraphics[scale=0.65,keepaspectratio]{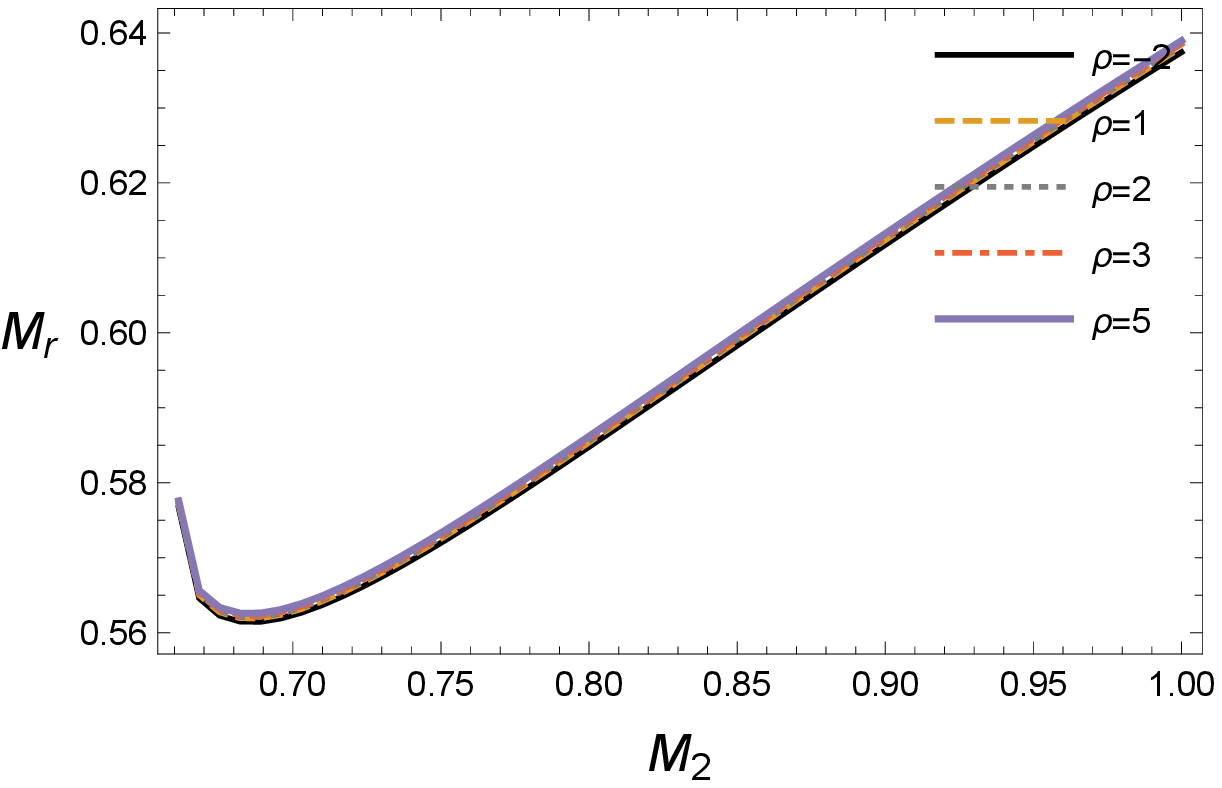}}
\caption{{\small The upper bound on the gravitational radiation with respect to $\rho$\,.}}
\label{fig:f121}
\end{figure}
Therefore, the parameter $\rho$ slightly affect the value of the radiation, but limits the range of the radiation by the extremal bound to be small, as shown in Fig.~\ref{fig:f91}~(b). In Fig.~\ref{fig:f121}~(a), the maximum radiation becomes small at the extremal values at smaller values of $\rho$, and the extremal bound on the spin parameter becomes smaller at larger values of $\rho$. For a given value of $\rho$, the radiation for large values of $\rho$ are  slightly greater than those for small values of $\rho$. The radiation about the change of the mass $M_2$ is almost not affected by the values of $\rho$, as can be seen in Fig.~\ref{fig:f121}~(b).

\begin{figure}[h]
\centering\subfigure[{{The upper bounds under $\rho=1$, $M_1=1$, $a_1=0.25$, and $a_2=0$\,.}}]
{\includegraphics[scale=0.9,keepaspectratio]{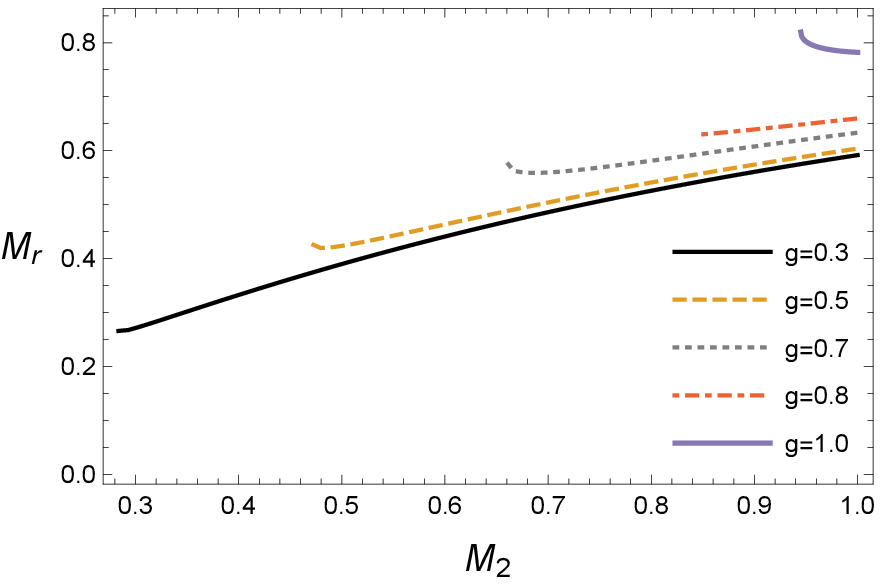}}
\quad
\centering\subfigure[{{The upper bounds under $\rho=-1$, $M_1=1$, $a_1=0.25$, and $a_2=0$\,.}}]
{\includegraphics[scale=0.9,keepaspectratio]{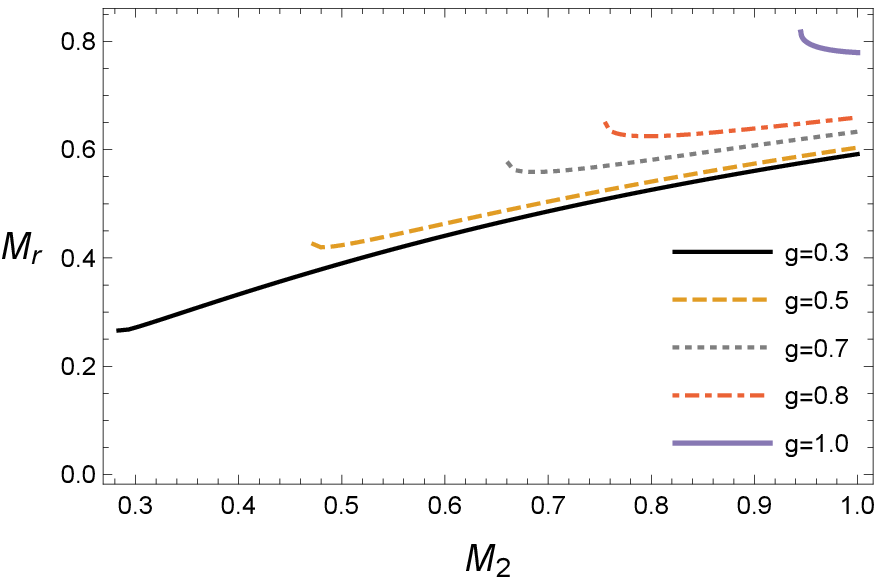}}
\caption{{\small The upper bound on the radiation with respect to $g$\,.}}
\label{fig:f171}
\end{figure}

\begin{figure}[h]
\centering\subfigure[{{The upper bounds under $\rho=1$, $M_1=1$, $a_1=0.15$, and $a_2=0$\,.}}]
{\includegraphics[scale=0.9,keepaspectratio]{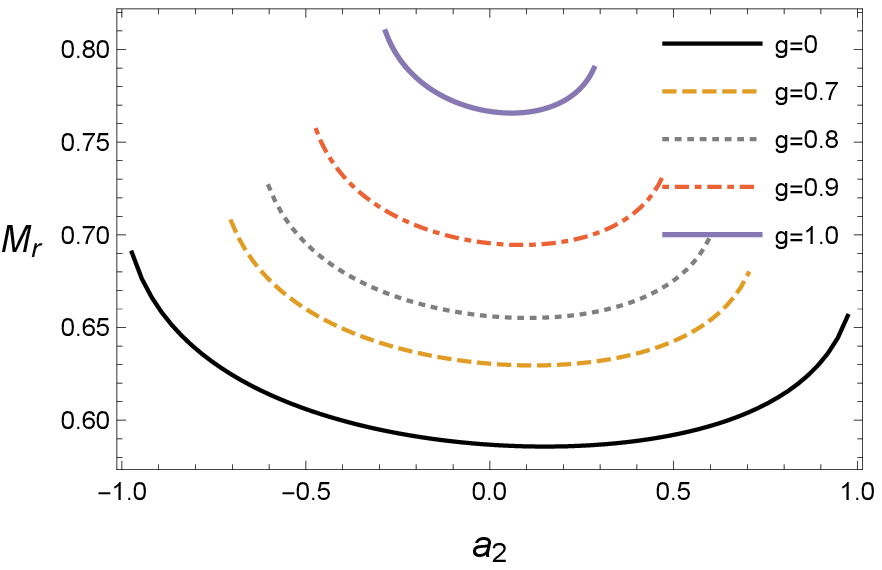}}
\quad
\centering\subfigure[{{The upper bounds $\rho=1$, $M_1=1$, $a_1=0.29$, $M_2$, and $a_2=0.15$\,.}}]
{\includegraphics[scale=0.9,keepaspectratio]{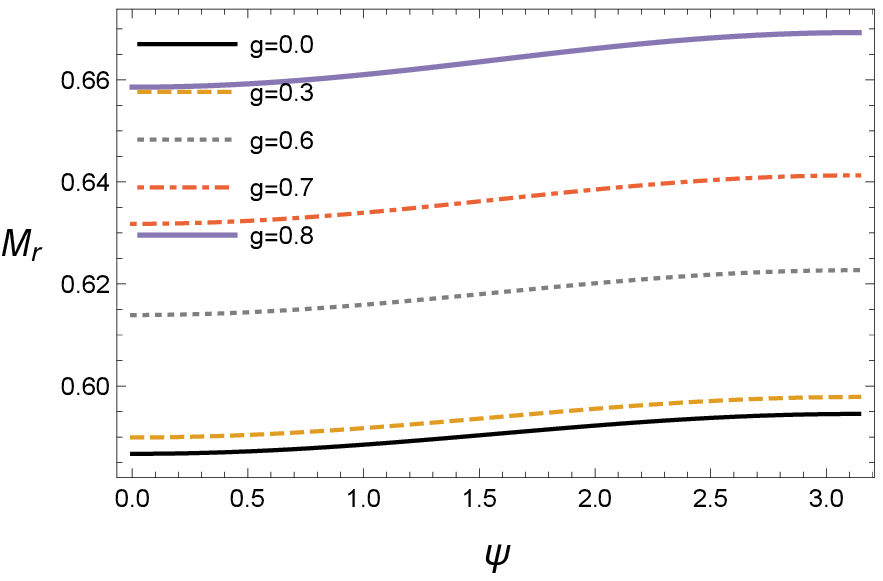}}
\caption{{\small The upper bound on the radiation with respect to $g$\,.}}
\label{fig:f151}
\end{figure}

The parameter $g$ is more important to the bound on the radiation than the parameter $\rho$. For a change of the mass $M_2$, the radiation is more sensitive to the parameter $g$ (Fig.~\ref{fig:f171}) than the parameter $\rho$ (Fig.~\ref{fig:f121}).
In addition, for the change in the parameter $g$, the response of the radiation to the sign of the parameter $\rho$ is almost negligible as seen in Fig.~\ref{fig:f171}~(a) and (b). The parameter $g$ is located in the mass function $m$ as a coefficient in the denominator, and is more affected by the mass and horizon than the parameter $\rho$. The parameter $g$ is also related to the extremal bound on the spin parameter $a$. The extremal bound becomes small at large values of the parameter $g$. Then, the minimum mass of the second black hole is limited, as shown in Fig.~\ref{fig:f171}. For a given mass, the radiation is released in a larger mass rather than a smaller mass. For the change of the spin parameter $a$, the overall behaviors are still similar to each values of $g$, but the amount of the radiation energy becomes large at large values of $g$, as shown in Fig.~\ref{fig:f151}.
More radiation is released in the anti-parallel alignment than in the parallel alignment, because the spin interaction is negative in the anti-parallel alignment, as shown in Fig.~\ref{fig:f151}~(a) and (b). Defined as a positive value, the parameter $g$ gives the minimum radiation at $g=0$, which represents the Kerr black hole. This can also be seen in Fig.~\ref{fig:f151}~(b), where the radiation depends on the angle $\psi$ and becomes large at large values of the parameter $g$.

\subsection{Choice of Parameters in Rotating Hayward Black Hole}
\begin{figure}[h]
\centering\subfigure[{The extremal spin parameter under $\rho=1$ and $M=1$\,.}]
{\includegraphics[scale=0.6,keepaspectratio]{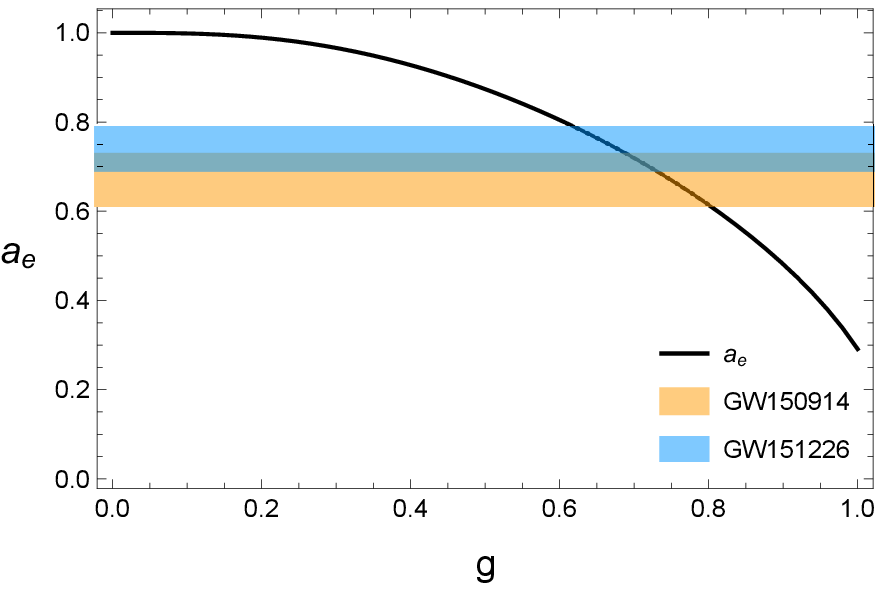}}
\quad
\centering\subfigure[{The minimum mass under $\rho=1$, $M=1$, and $a=0$\,.}]
{\includegraphics[scale=0.6,keepaspectratio]{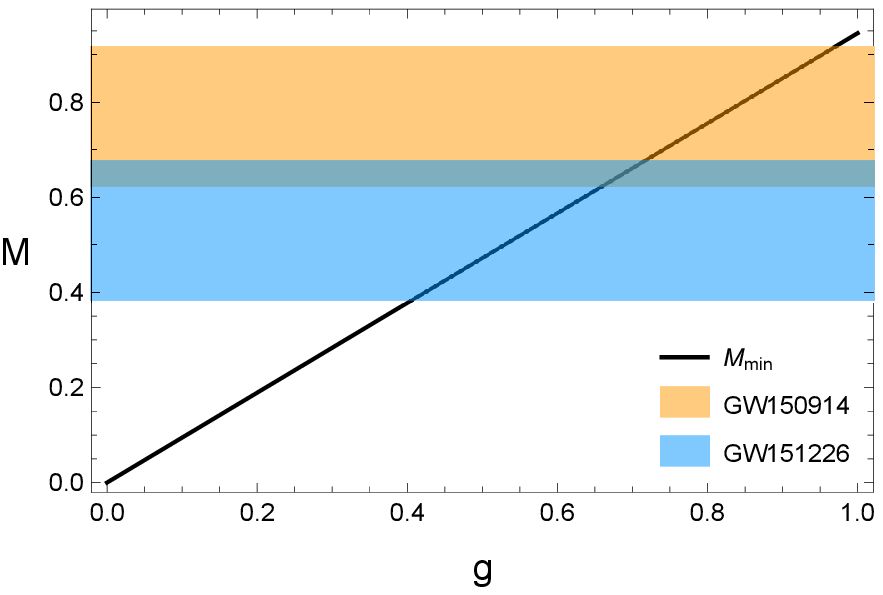}}
\quad
\centering\subfigure[{The upper bounds on the radiation for $\rho=1$, $M_1=1$, $a_1=0$, and $a_2=0$\,, and the slope between LIGO data\,.}]
{\includegraphics[scale=0.6,keepaspectratio]{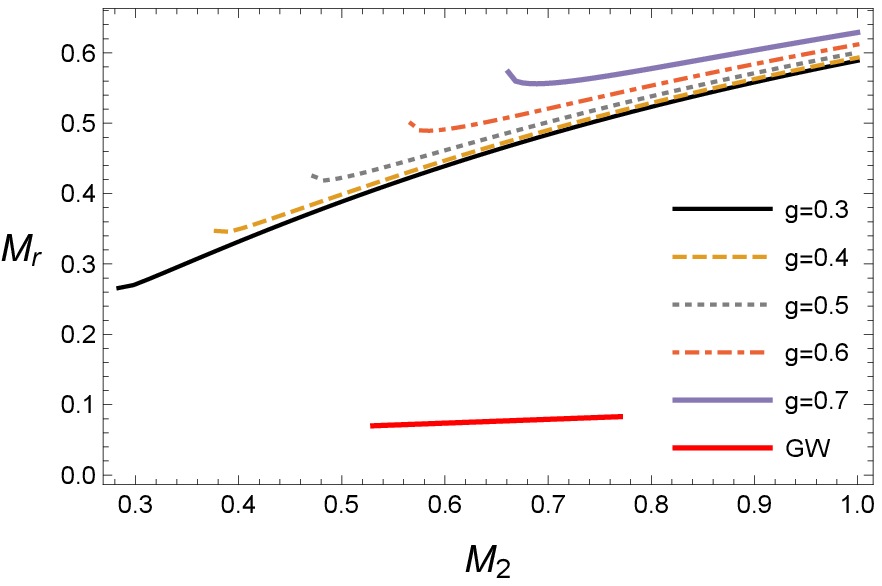}}
\caption{{\small The upper bounds thermally allowed and detections of the LIGO\,.}}
\label{fig:f501}
\end{figure}

We will approximately apply our results and the properties of the Hayward black hole to the GW150914 and GW151226 detected at LIGO to find the bound of the parameter $g$. In this work, we set the most massive black hole in the initial state to unity, and the black holes are described in consideration of the mass ratio. The detections are compared with respect to the final spin parameter $a_3$, minimum mass $M_{min}$, and radiation slope $M_r$ in Fig.~\ref{fig:f501}. Note that the data of the GW150914 and GW151226 have model dependencies and include errors\cite{Abbott:2016blz,Abbott:2016nmj,Ghosh:2015jra,TheLIGOScientific:2016wfe}, but our approach is not to fix the exact value of the parameter, and hence it will not change the result presented in this section.
In Fig.~\ref{fig:f501}~(a), the spin parameters of the final black holes, $0.67^{+0.06}_{-0.08}$ and $0.74^{+0.06}_{-0.06}$ in the GW150914 and GW151226 are given with the extremal spin parameters for each values of $g$. As the value of $g$ becomes large, the extremal spin parameter decreases, so that the spin parameter of the final black hole should be included in the extremal value at least. As shown for the spin parameters with their range in Fig.~\ref{fig:f501}~(a), too great a large value of $g$ is not allowed. The possible upper bound of the parameter $g$ ranges between $0.7$ and $0.8$, and the value of the overlap range is at $g=0.7$. In Fig.~\ref{fig:f501}~(b), the mass ratios of the second black hole are given with their ranges. In the setting $M_1$ to unity, the second black hole is smaller than the first, and the mass of the second black hole should be inside of the minimum mass for a given value of $g$. The mass ratios, estimated $0.79^{+0.18}_{-0.19}$ and $0.53^{+0.31}_{-0.31}$ in the GW150914 and GW151226\cite{Abbott:2016blz,Abbott:2016nmj,TheLIGOScientific:2016wfe} are applied approximately with the minimum mass in Fig.~\ref{fig:f501}~(b). The ratio ranges are large, and hence a possible value of $g$ is also large and overlaps at approximately $g=0.7$, interestingly. We use the relation of the gravitational radiations between GW150914 and GW151226 in Fig.~\ref{fig:f501}~(c). The energy ratios radiated in the GW150914 and GW151226 are about $0.07$ and $0.083$. Their slope of $M_{r,bound}$ is from $7$ to $8$ times of the data, but the order of times is the same as the ratio between $M_{r,bound}$ and $M_r$, so that it is not from the slope itself. As shown in Fig.~\ref{fig:f501}~(c), the slope becomes low at the large value of $g$ allowed by the minimum mass. Using Fig.~\ref{fig:f501}~(c), the allowed value of $g$ might be smaller than $0.5$. Therefore, we can find out that the value of $g$ may not become too large a value in the regular black hole model. However, whatever our results are in this section, more detection is needed to determine the precise value of the parameter $g$.

\section{Summary and Conclusion}\label{sec5}
We investigated the spin interaction and gravitational radiation released in the collision of two Hayward black holes. The Hayward black hole is constructed from a modified Einstein equation, having no curvature singularity in the whole spacetime due to the regularity effects. To find out the extent of influence of the effect, indicated by $g$ and $\rho$, on the radiation released in the collision, we focused on the spin interaction as well as the thermally allowed upper bound on the radiation. We supposed an initial state where two Hayward black holes stay far from each other and have an angle difference of $\psi$ between two angular momenta of the black holes. Slowly coming together, the two black holes endure a head-on collision and become a Hayward black hole in the final state. In this procedure, we imposed the angular momentum conservation and the second law of thermodynamics. Then, according to the first law of thermodynamics, the loss of the mass in the final state was equated to the gravitational radiation released in the collision.

First, the potential of the spin interaction was exactly obtained using the MPD equation, when one of black holes was approximated to a spinning particle as a limit of the small mass. Then, we compared the potential with the radiation in the limit of the small mass and parameter $g$. The potential and radiation corresponded in the limit of $g$ going to zero, the Kerr black hole case, but they did not at the value of $g$ greater than zero. However, we found out that the alignment of angular momenta between black holes surely affects the radiation due to the potential of the spin interaction for any values of the parameter $g$. The potential becomes positive in the parallel alignment $\psi=0$, and the released radiation is smaller by as much as the value of the potential. On the contrary, the anti-parallel alignment $\psi=\pi$ has the negative potential, and hence more radiation can be released, that is, as much as the interaction energy.

We obtained numerically the thermally allowed upper bound on the radiation. As expected in the analytical approach, the radiation depended on the alignment of the angular momenta, denoted as $\psi$, because of the contribution of the spin interaction. Then, the anti-parallel case released greater energy than the parallel case. The bound of the radiation energy responded more sensitively to the parameter $g$ than to $\rho$. As the parameter $\rho$ increased from the negative value, the bound of the radiation also increased and ended at the finite positive value owing to the extremal condition, but the increase of the radiation was very small compared with that of $\rho$. For the increase of the parameter $g$, the bound on the radiation also increased, and the range of the spin parameter allowed in the extremal condition became narrow. Then, we fixed the value of $\rho$ to unity and determined the range of $g$ value using the GW150914 and GW151226 detected by LIGO. Actually, our estimation of the value of $g$ was approximate, and the range of $g$ is still broad. However, the possible value of $g$ is estimated to be less than $0.7$ when using the extremal spin parameter and minimum mass of the Hayward black hole applied to the final state of the black hole in the binary black hole merger. In addition, we used the slope of the radiation with respect to the mass ratios of the GW150914 and GW151226. Considering the minimum mass and the order of difference between the bound and radiation, the value of $g$ is now expected to be smaller than $0.5$. To improve our estimation, we need to detect more gravitational wave generated by a black hole binary having a very small mass ratio, because our analysis becomes more precise in the limit at which one of the black holes has a very small mass compared with the other.
\\

{\bf{Acknowledgments}}

{\small This work was supported by the faculty research fund of Sejong University in 2016. BG was supported by Basic Science Research Program through the National Research Foundation of Korea(NRF) funded by the Ministry of Science, ICT \& Future Planning(NRF-2015R1C1A1A02037523).}

\begin{appendices}
\section{Hayward Black Hole in Nonlinear Electrodynamics}
The Hayward black hole is not only a solution of a modified Einstein equation\cite{Hayward:2005gi}, but is also found in the Einstein gravity coupled with nonlinear electrodynamics\cite{Fan:2016hvf}. In this appendix, we briefly review the Hayward black hole in nonlinear electrodynamics and show the bounds on the gravitational radiation. In fact, other regular black holes were obtained in the Einstein gravity coupled with nonlinear electrodynamics before the Hayward black hole\cite{AyonBeato:1998ub,AyonBeato:1999ec,AyonBeato:1999rg,AyonBeato:2000zs,Junior:2015fya,Fan:2016rih,Fan:2016hvf}. The action is given as
\begin{eqnarray}\label{eq:nonlinear1}
S=\frac{1}{16\pi}\int d^4 x \sqrt{-g}\left(R-\mathcal{L}(\mathcal{F})\right)\,, \quad \mathcal{F}=F^{\mu\nu}F_{\mu\nu}\,,\quad F=dA\,,
\end{eqnarray}
where $F$ is the Maxwell field strength. The equations of motion are obtained from Eq.~(\ref{eq:nonlinear1}) as
\begin{eqnarray}\label{eq:nonlinear2}
R_{\mu\nu}-\frac{1}{2}g_{\mu\nu}R=2\left(\mathcal{L}_{\mathcal{F}}F_{\mu\nu}^2-\frac{1}{4}g_{\mu\nu}\mathcal{L}\right)\,,\quad \nabla_\mu \left(\mathcal{L}_{\mathcal{F}F^{\mu\nu}}\right)=0\,, \quad \mathcal{L}_{\mathcal{F}}=\frac{\partial \mathcal{L}}{\partial \mathcal{F}}\,.
\end{eqnarray} 
Under the equations of motion, we consider the ansatz of the static solution given as
\begin{eqnarray}\label{eq:nonlinear3}
ds^2=-f(r)dt^2+\frac{1}{f(r)}dr^2+r^2(d\theta^2+\sin^2\theta d\phi^2)\,, \quad A=Q_m \cos\theta d\phi\,,
\end{eqnarray}
where $Q_m$ is the magnetic charge related to the value of $g$. The Hayward black hole appears in a large class of the solution given in Eq.~(\ref{eq:nonlinear3}) applied to Eq.~(\ref{eq:nonlinear2}). When we specify the Lagrangian density $\mathcal{L}(\mathcal{F})$ to
\begin{eqnarray}\label{eq:nonlinear4}
\mathcal{L}(\mathcal{F})=\frac{12\left(\sigma \mathcal{F}\right)^{3/2}}{\sigma\left(1+\left(\sigma\mathcal{F}\right)^{3/4}\right)^2}\,,
\end{eqnarray}
where the mass $M_g$ is the magnetic charge that differs from the Schwarzschild mass. The solution of Eq.~(\ref{eq:nonlinear4}) is obtained as
\begin{eqnarray}\label{eq:nonlinear5}
f(r)=1-\frac{2M_g r^2}{r^3+g^3}\,,\quad M_g=\frac{g^3}{\sigma}\,,
\end{eqnarray}
which is the static Hayward black hole, so that we can obtain the rotating Hayward black hole having a mass $M_g=M$. In nonlinear electrodynamics, the reason for regularity is in the mass $M_g$, which comes from the magnetic charge $Q_m$. Regular black holes in nonlinear electrodynamics have zero Schwarzchild mass, which generates the curvature singularity. The magnetic charge $Q_m$ is given in terms of $g$ and $\sigma$
\begin{eqnarray}\label{eq:nonlinear6}
Q_m=\frac{g^2}{\sqrt{2\sigma}}\,,
\end{eqnarray}
where the value of $g$ is limited to a positive number, because the black hole solution only exists at the positive value. This is a very brief review of the Hayward black hole in nonlinear electrodynamics. The detailed properties can be found in Ref.\cite{Fan:2016hvf}\,.

\begin{figure}[h]
\centering\subfigure[{The upper bounds under $\rho=1$, $g_1=0.2$, $M_1=1$, $a_1=0.2$, and $g_2=0.2$\,.}]
{\includegraphics[scale=0.9,keepaspectratio]{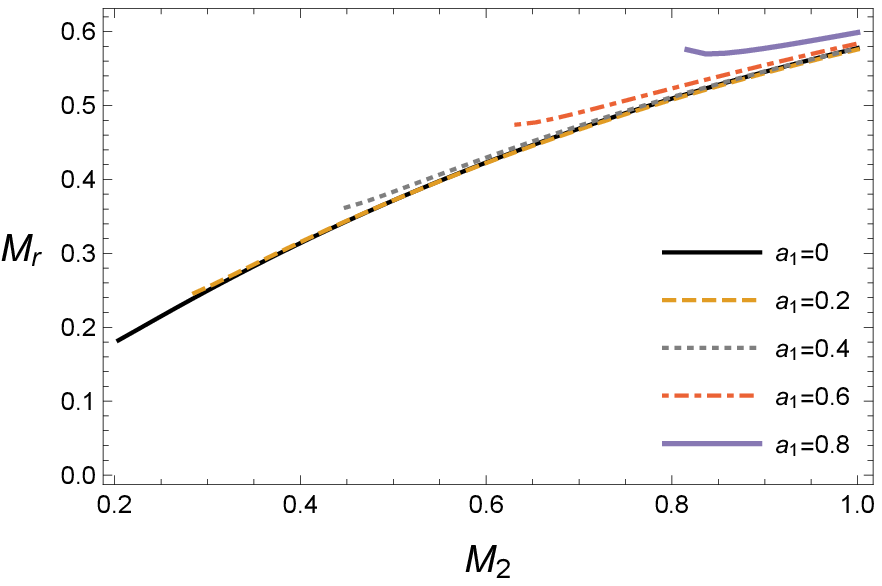}}
\quad
\centering\subfigure[{The upper bounds under $\rho=1$, $g_1=0.2$, $M_1=1$, $g_2=0.2$, and $M_2=1$\,.}]
{\includegraphics[scale=0.9,keepaspectratio]{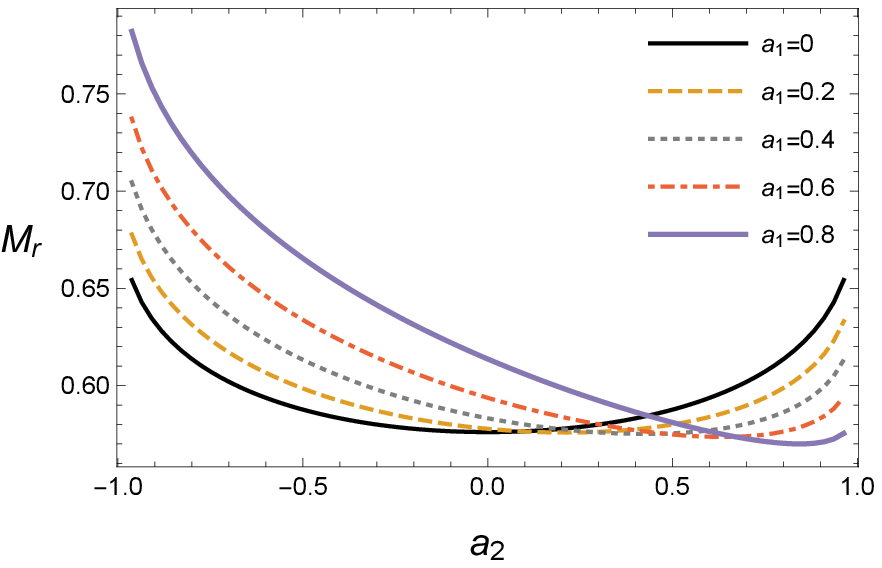}}
\caption{{\small The upper bounds on the radiation in nonlinear electrodynamics\,.}}
\label{fig:fA1}
\end{figure}

\begin{figure}[h]
\centering\subfigure[{The upper bounds under $M_1=1$, $a_1=0.2$, $g_2=0.2$, $M_2=1$, and $a_2=0.2$\,.}]
{\includegraphics[scale=0.6,keepaspectratio]{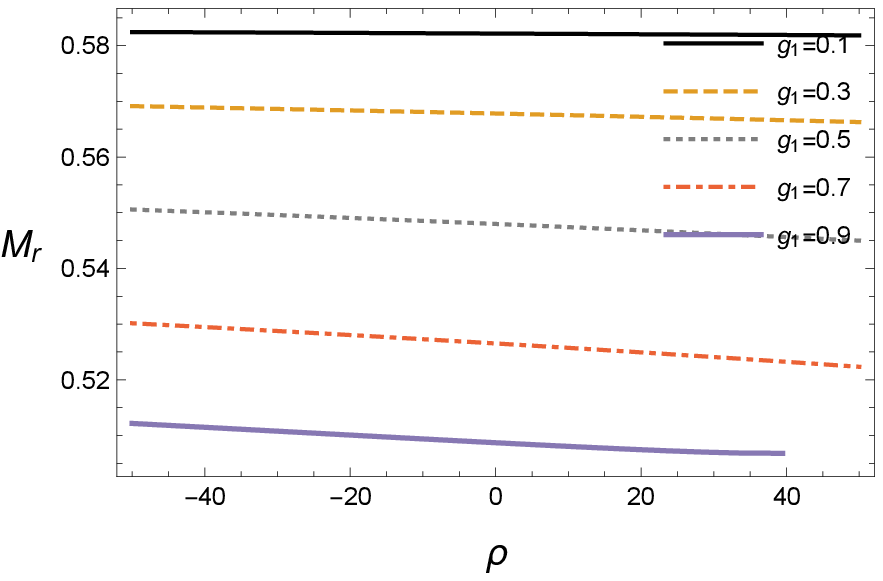}}
\quad
\centering\subfigure[{The upper bounds for $g_1=g_2$ cases under $\rho=1$, $M_1=1$, $a_1=0.2$, and $M_2=1$\,.}]
{\includegraphics[scale=0.6,keepaspectratio]{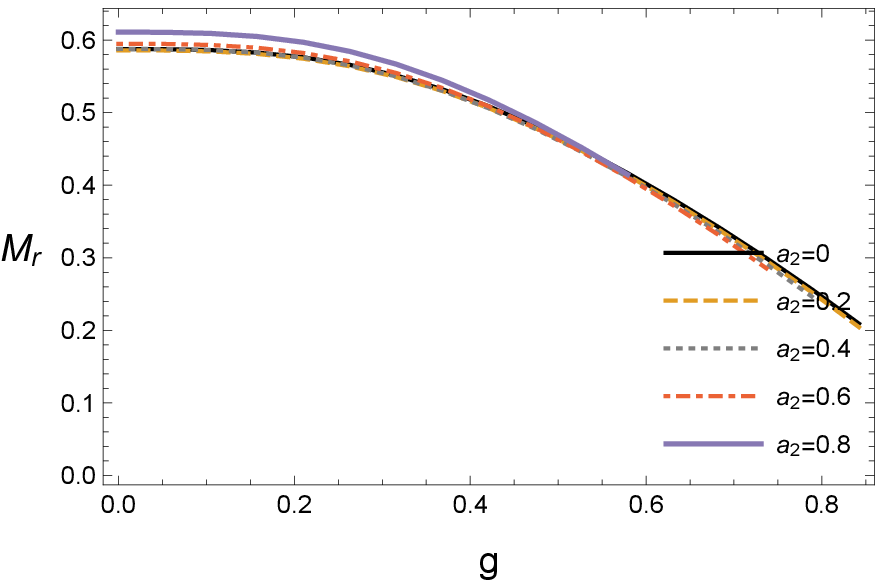}}
\quad
\centering\subfigure[{The upper bounds for $g_1\not= g_2$ under $\rho=1$, $M_1=1$, $a_1=0.2$, $M_2=1$, and $a_2=0.2$\,.}]
{\includegraphics[scale=0.6,keepaspectratio]{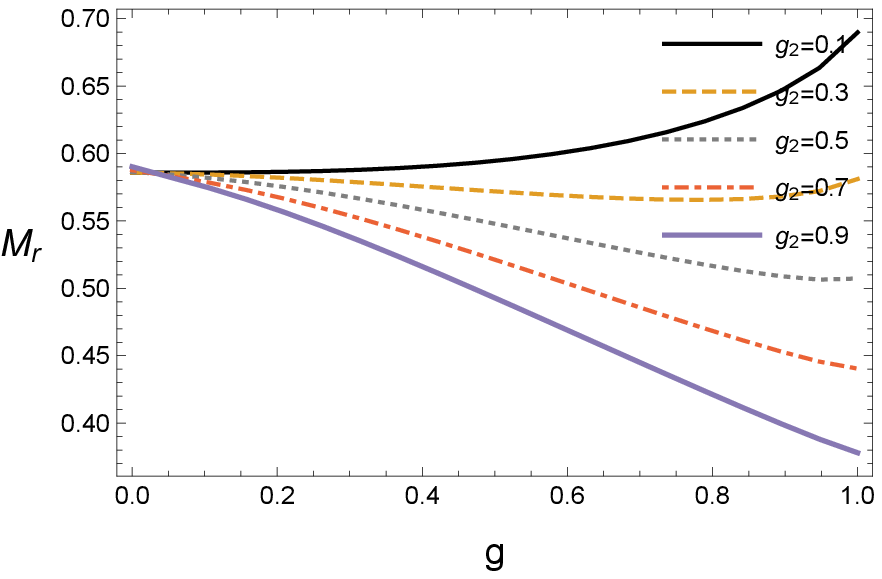}}
\caption{{\small The upper bounds on the radiation with respect to $\rho$ and $g$\,.}}
\label{fig:fA2}
\end{figure}

We also simply introduce the bounds on the radiation for the Hayward black hole in nonlinear electrodynamics. The only difference regarding the setting of the parameter $g$ is related to the magnetic charge. Following the analogy in section~\ref{sec4}, we assume the collision of two black holes having different magnetic charges in the initial state: one is the mass $M_1$ and magnetic charge $Q_1$; the other is $M_2$ and $Q_2$. In the final state, the merged black hole will be $M_3$ and $Q_3$. Other conditions are the same as in section~\ref{sec4}, and we impose only one more condition about the conservation of the magnetic charge, which is rewritten as
\begin{eqnarray}
Q_1+Q_2 = Q_3\,,\quad \sqrt{M_1g_1}+\sqrt{M_2g_2}=\sqrt{M_3g_3}\,,
\end{eqnarray}
where the rewritten form of the conservation is from Eq.~(\ref{eq:nonlinear5}) and (\ref{eq:nonlinear6}). Then, the bounds on the gravitational radiation are given in Fig.~\ref{fig:fA1}.
The overall behaviors of the upper bounds are similar to those of a modified Einstein equation in Fig.~\ref{fig:f7}. The radiation increases as the mass of black holes increases. In addition, the alignment of black holes contributes to the radiation in the same manner as does a Hayward black hole in a modified Einstein equation. The upper bounds are also dependent on the parameters $\rho$ and $g$, as shown in Fig.~\ref{fig:fA2}. The parameter $\rho$ becomes large, and then the bound on the radiation decreases, as shown in Fig.~\ref{fig:f91} and \ref{fig:fA2}~(a). However, the response to the change of the parameter $g$ is in Fig.~\ref{fig:fA2}~(b) and (c), and is contrary to the case shown in Fig.~\ref{fig:f171} and \ref{fig:f151}. Since the area of the horizon becomes small at a large value of $g$, an increase of $g$ reduces the area of the final black hole according to the conservation of the magnetic charge, so that the mass of the final black hole becomes larger than that of the modified Einstein equation. Therefore, in nonlinear electrodynamics, the radiation of the Kerr black hole case is the largest in varying parameter $g$\,.
\end{appendices}

\end{document}